\documentclass[useAMS,usenatbib]{mn2e}
\usepackage{graphicx,amssymb,url,enumerate}
\usepackage{epstopdf}

\def\lesssim{\la}
\def\kmsp{km~s$^{-1}$~kpc}

\title[Parent populations of abundance groups]{The Parent Populations of 6 groups identified from Chemical Tagging in the Solar neighborhood}
\author[Quillen et al.]{
Alice C. Quillen$^1$, 
Borja Anguiano$^{2,3}$, 
Gayandhi De Silva$^{4,5}$, 
Ken Freeman$^6$,
\newauthor 
Dan B. Zucker$^{2,3,4}$,  
Ivan Minchev$^7$ and
Joss Bland-Hawthorn$^5$
\\
$^1$ Department of Physics and Astronomy, University of Rochester, Rochester, NY 14627, USA\\
$^2$ Macquarie University Research Centre in Astronomy, Astrophysics \& Astrophotonics, NSW 2109,  Australia \\
$^3$ Department of Physics \& Astronomy, Macquarie University, NSW 2109, Australia \\
$^4$ Australian Astronomical Observatory, PO Box 296, NSW 1710, Australia\\
$^5$ Sydney Institute for Astronomy, School of Physics, University of Sydney, NSW 2006, Australia\\
$^6$ Research School of Astronomy \& Astrophysics, Mount Stromlo Observatory, ACT 2611, Australia \\
$^7$ Leibniz-Institut f\"ur Astrophysik Potsdam (AIP), An der Sternwarte 16, 14482, Potsdam, Germany \\
}

\begin{document}
\maketitle

\begin{abstract} 
We estimate the size and distribution of the parent populations for
the 6 largest (at least 20 stars in the Solar neighborhood) 
chemical groups identified in the Chemical Tagging experiment by Mitschang et al.~2014.
Stars in the abundance groups tend to lie near a boundary in angular momentum versus eccentricity space 
where the probability is highest for a
star to be found in the Solar neighborhood and  where orbits have apocenter approximately equal to the Sun's galactocentric radius.
Assuming that the parent populations are uniformly distributed at all azimuthal angles in the Galaxy,
we estimate that the parent populations of these abundance groups 
contain at least 200,000 members.  
The spread in angular momentum of the groups implies that the assumption of a uniform azimuthal distribution 
only fails for the two youngest groups and only for the highest angular momentum stars in them. 
The parent populations of three thin disk groups have narrow angular momentum distributions, but
tails in the eccentricity and angular momentum distributions suggest that only a small fraction of stars have migrated
and increased in eccentricity.
In contrast, the parent populations of the thick disk groups exhibit both wide angular momentum and eccentricity distributions
implying that both heating and radial migration has taken place.

\end{abstract}

\section{Introduction}

\citet{freeman02} proposed that stars with similar abundance measurements could represent a particular star formation
or enrichment event, discrete in space and time, as might be expected from the homogeneity of nearby
open clusters and moving groups \citep{desilva06, desilva07, quillen02}.  These stars 
 would subsequently disperse in the Galaxy, retaining their initial chemical patterns \citep{bland10}.
A search for stars that have a similar abundance pattern as the Sun would allow us to learn about the
birth place of the Sun \citep{simon09,liu15}.
Because stars in the solar neighborhood span a wide distribution in stellar ages, metallicities and inferred
birth Galactocentric radii, it is difficult to pin down the role of specific mechanisms for stellar migration 
and heating (increase in radial and vertical epicyclic motions) 
(e.g., see \citealt{quillen09,schoenrich09,freeman02,kruijssen11,haywood13,minchev14,lehnert14}).   
A study of homogeneous groups of stars should give complementary constraints
on migration and heating processes, compared to those arising from studies of 
 heterogeneous distributions (such as a magnitude limited sample of stars in the Solar neighborhood).   
 
We focus on the a high-resolution spectroscopic study of 714 F and G dwarf and subgiant stars in the Solar neighborhood
studied by \citet{bensby14}.
The blind chemical tagging experiment by \citet{mitschang14} used this sample to identify groupings
  of nearby disc field stars 
that share  metal abundance measurements.  The field stars they identified as having similar abundances are not clustered
in space, nor do they share similar space motions.  
Using  isochrone sets, \citet{mitschang14} estimated the ages of each of these chemical groupings. 
These groups represent a first attempt to identify  groups of stars from single discrete birth events.

We ask here:  what is the number and distribution in the Galaxy of a parent stellar population of one of these abundance groups? 
We necessarily focus on only the 6 largest groups identified by \citet{mitschang14} each of which contains more than 20 stars.
We begin by assuming that the parent population for each group is currently evenly distributed (azimuthally) in the Galaxy 
and at the current time only a fraction of the stars in the parent population are present in the Solar neighborhood. 
This assumption was adopted for the toroid models by \citet{bland10} (for an illustration see their Figure 4). 
This assumption neglects how a cluster
 dissolves and is  dispersed in the Galaxy (see discussions by \citealt{bland10,simon09}).
We will discuss how our assumption of axi-symmetry for the parent distribution
 could have impacted our inferred parent population distributions.
 We also neglect our location in the Galaxy with respect to  spiral and bar perturbations, 
(e.g.,  \citealt{quillen11,quillen14,minchev13,minchev14}). We  
assume there is no correlation between vertical oscillation amplitude
 and eccentricity in the parent population and that the epicyclic angle distribution is relaxed 
 (see \citealt{minchev09} for an illustration of what can be seen when this is not true).

We first consider the fraction of time that a star with a given eccentricity 
and angular momentum
might be seen in the Solar neighborhood.\footnote{We 
only consider the $z$ component of angular momentum that is dominated by rotation in the Galaxy.
In the Solar neighborhood the angular momentum of a star 
$L \approx R_\odot (V + V_{LSR,\odot})$ with $R_\odot$ the galactocentric 
radius of the Sun, $V_{LSR}$ the rotation velocity of the local standard of rest 
and $V$ the tangential component of the star's velocity vector.}
From a distribution of orbits with a given eccentricity and angular momentum 
 we estimate the probability that a star is seen in a Solar neighborhood volume
with boundary dependent on the distances of stars in the \citet{bensby14} sample.    
For each star in one of Mitschang et al.'s chemical groupings the inverse of this probability  
lets us estimate  the number of stars at similar eccentricities and angular momentum in the parent population.

 \citet{bensby14} selected stars for spectroscopic study  with a range of properties and necessarily did not
observe every F and G star in the Solar neighborhood.
We compare the \citet{bensby14} sample with 
 the Geneva-Copenhagen survey of F and G stars in the Solar neighborhood \citep{nordstrom04,holmberg09}
to estimate a selection bias as a function of angular momentum.
The GCS is a magnitude limited, kinematically unbiased sample of almost 17000 nearby F and G stars. 
\citet{bensby14} warn that their sample is a compilation of a number
of different observing programs and so they give no selection description for the entire sample.
We should be careful interpreting inferred parent distributions, keeping in mind that there
might be additional biases arising from the selection of this sample.

Using these two corrections, the first based on probability for such an orbit to be seen in the Solar neighborhood, the second
based on selection bias,
we derive estimates for the source or parent populations of the 6  abundance groups  identified by \citet{mitschang14}.
A discussion follows on the nature of the parent populations and on how our underlying assumptions 
have impacted our estimate of their number and distributions.

\begin{figure*}
\includegraphics[width=5.0in, trim= 0 0 0 0 ]{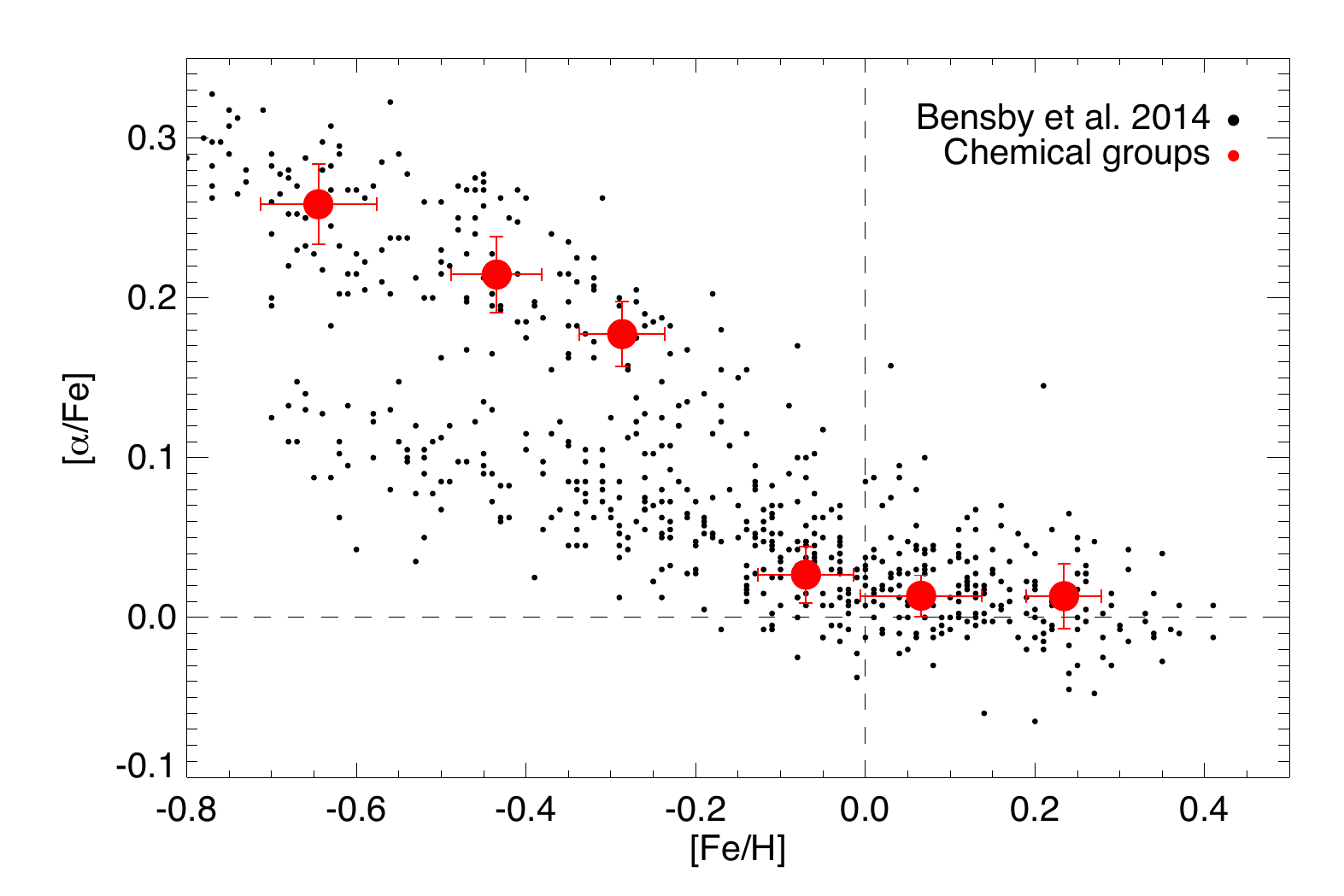}
\caption{
Mean abundances (in solar units) for each of the six abundance groups (found by \citealt{mitschang14}) 
are shown as large red dots.  
Error bars represent standard deviations of the abundance values of stars in the group.
Black dots show abundances
of individual stars from the \citet{bensby14} sample.
The dashed lines show solar values.  The three youngest 
groups are typical of the thin disk population, whereas the three oldest are typical of the thick disk population.  
\label{fig:fealpha} } 
\end{figure*}

\begin{figure*}
\includegraphics[width=4.0in, trim= 0.5truein 0 0 0 ]{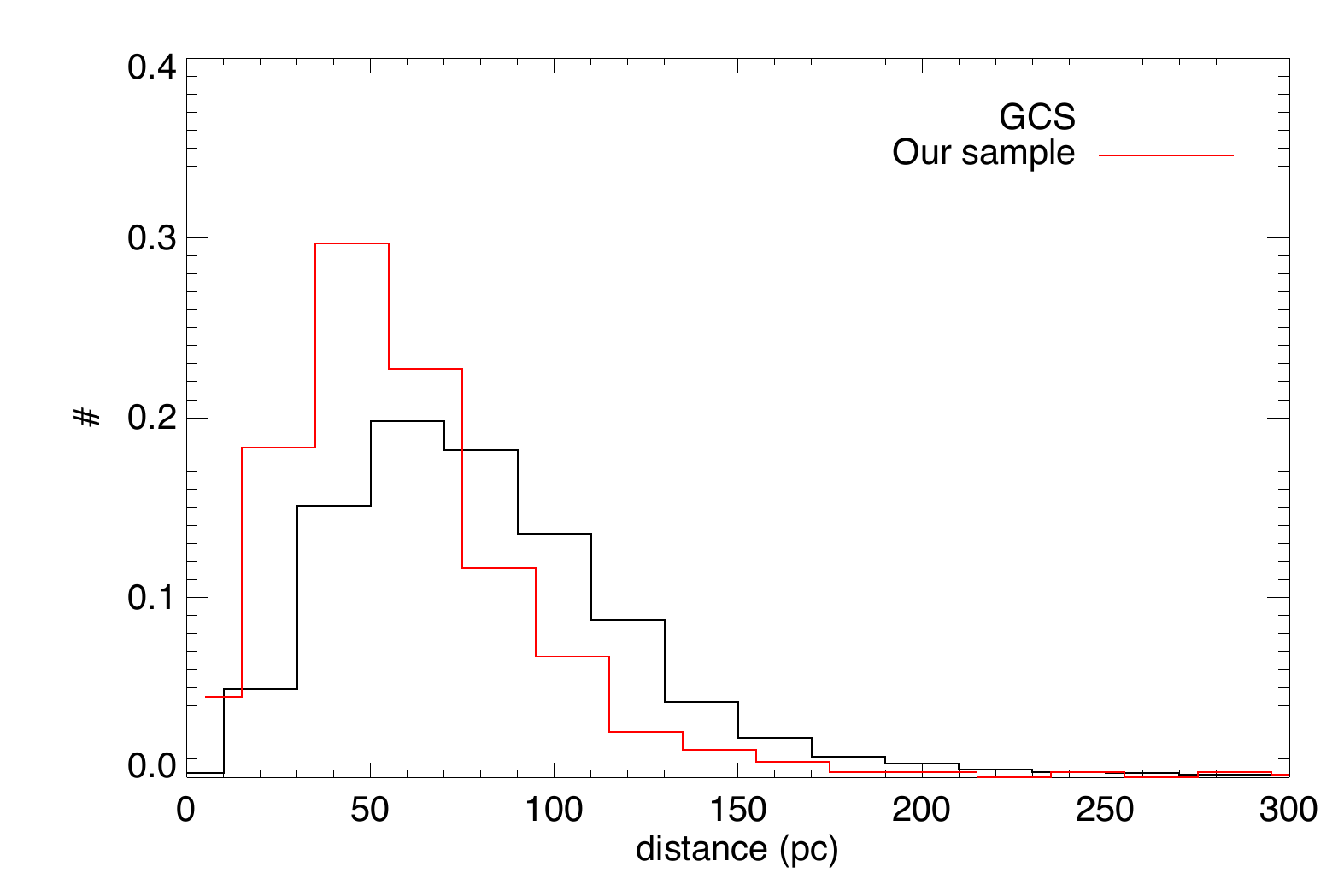}
\caption{Distance distribution for the \citet{bensby14} sample compared to that of the Geneva-Copenhagen Survey \citep{nordstrom04,holmberg09}.
Most stars from the \citet{bensby14} sample are within 100 pc of the Sun.  
The distributions have been normalized so that they integrate to 1. 
\label{fig:distances}  } 
\end{figure*}

\begin{table*}
\begin{minipage}{100mm}
 \caption[]{Properties of the Abundance Groups \label{tab:groupings}}
 \begin{tabular}{l r r  r r r r r r }
 \hline
 GID & N  & Age & [Fe/H] & [$\alpha$/Fe]  & $\langle L \rangle$ &$\sigma_L$ & $\langle e \rangle$  \\
         &     & (Gyr) &           &                      &   \multicolumn{2}{c}{(\kmsp)}         &                 \\
 \hline
5 &21 & $4.0\pm 0.6$     & $0.23\pm0.04$ & $0.01\pm 0.02$ & 1715 & 214 & 0.14\\ 
1 &42 & $4.8\pm 0.4$    & $0.07\pm 0.07$ & $0.01\pm 0.01$ & 1748 & 262 & 0.17 \\
3 &25  & $7.1\pm 0.4$   & $-0.07\pm 0.06$ & $0.03\pm 0.02$ & 1522 & 247 & 0.28\\
4 &24  & $10.1\pm 1.4$  & $-0.43\pm0.05 $& $0.18\pm 0.02$ & 1450 & 271 & 0.32 \\
2 &30 & $10.2\pm 0.8$  & $-0.30\pm 0.05$ & $0.22\pm 0.02$ & 1368 & 373 & 0.34\\
6&21 & $12.1\pm 1.1$  & $-0.64\pm 0.07$ & $0.26\pm 0.02$ & 1297 &  510 & 0.39 \\ 
 \hline
\end{tabular}
 \\
GIC is the group number given by \citet{mitschang14}.  
$N$ is the number of stars in the abundance group and the estimated age by \citet{mitschang14} is given in Gyr.
Mean [Fe/H] and [$\alpha$/Fe] values in solar units for the stars in the group were computed 
using abundance values listed in Table C.3 by \citet{bensby14}, finding them using Hipparcos catalog 
numbers for the stars listed in Table 2 by \citet{mitschang14} for each  group.  
Errors in the abundances are the standard deviations of the abundance values from each star.
We list the mean $\langle L\rangle $ and standard deviation, $ \sigma_L$, (in \kmsp)
 of the angular momentum distributions and the mean eccentricity $\langle e \rangle $
 for each group (distributions are shown in Figure \ref{fig:hist}).  These are  computed from
 the eccentricity and angular momentum values computed by \citet{bensby14}.
 \end{minipage}
\end{table*}

\section{Properties of the six abundance groups }
\label{s:groups}

Properties of the 6 largest abundance groups found by \citet{mitschang14} are listed in Table \ref{tab:groupings}.
The  Hipparcos catalog numbers of the stars in each group
are listed in Table 2 by \citet{mitschang14}.
Our Table \ref{tab:groupings}  lists the group identification number (from their Table 2), 
group age (that derived by \citealt{mitschang14} using
Yonsei-Yale isochrone sets and with error estimate described in their section 4.2)
the mean [Fe/H] of the group (in Solar units and using abundances listed in Table C2 by \citealt{bensby14}) and the group mean 
[$\alpha$/Fe] in Solar units.  
The mean abundance values and standard deviation for each group are computed
 from the values for each star in the group. 
For each star  [$\alpha$/Fe] is calculated by averaging the abundances 
for $\alpha$ elements Ti, Mg, Si, and Ca (as done by \citealt{mitschang14}).
Table \ref{tab:groupings} also lists
the mean $\langle L\rangle $ and standard deviation, $ \sigma_L$, (in \kmsp)
 of the angular momentum distributions of each group.  These are  
 the angular momentum values  by \citet{bensby14} who
  computed space motions for all the stars in their sample 
(see their section 3).\footnote{The adopted local standard of rest
($U_\odot,V_\odot, W_\odot$) = (11.10, 12.24, 7.25) km~s$^{-1}$  is that
by \citet{schoenrich10}.}  

Studies of abundance populations based on high resolution spectroscopy find a bi-modality in the abundance distribution 
(e.g., \citealt{navarro11,fuhrmann11,haywood13,adibekyan13,anders14} and references therein)
with a dividing line between thin and thick disks populations
near [$\alpha$/Fe]$\sim 0.12$  (e.g., see Figure 12 by \citealt{reddy06} and section 5.1 by \citealt{mitschang14}).
For each abundance group,  mean values of [Fe/H] are plotted against the mean values of [$\alpha$/Fe] in Figure \ref{fig:fealpha}
with the other stars in the \citet{bensby14} sample.
The three youngest groups have  abundances consistent with a thin disk population,
whereas the older three have abundances consistent with a thick disk population.
 
690 out of 714 stars in the \citet{bensby14} sample are also present in the Geneva-Copenhagen Survey of F and G stars in the solar neighborhood  (GCS) \citep{nordstrom04,holmberg09}.  
The distribution of distances from the \citet{bensby14} sample is compared to that of the GCS 
in Figure \ref{fig:distances}, illustrating that the stars in the \citet{bensby14} sample are predominantly nearer than 100 pc.
Here distances are based on parallaxes from the new reduction of the Hipparcos data by \citet{van07}.
Of the 163 stars in the 6 abundance groups, we find that
only 9 of the stars are further than 100 pc from the Sun.  
Thus this sample of stars is confined to a small spherical volume, centered on the Sun, 
with an approximate radius of 100 pc.  

From the angular momentum, $L$, and eccentricity, $e$, values listed by \citet{bensby14} we constructed
histograms for each group, and these are shown in Figure \ref{fig:hist}.
The mean angular momentum (also listed in Table \ref{tab:groupings}) 
for each group decreases with increasing age, suggesting that
the oldest groups arise from the inner galaxy and the youngest groups are located near the Sun's galactocentric radius. 

While \citet{bensby14} did not list errors for eccentricity $e$ or angular momentum $L$ for each star,
 we can assume that the space velocity components  $U,V,W$
have errors the same size as those of the GCS which are estimated to be $\Delta v~\sim 1.5$ km~s$^{-1}$ 
(see section 4.7 by \citealt{nordstrom04}).
This corresponds to an approximate error of $\Delta L \sim 13$ \kmsp ~ in angular momentum.
We estimate the size of an error in eccentricity with $\Delta e \sim \Delta v/V_{LSR,\odot} \sim 0.01$. 
 The errors could also have
systematic trends in them (as a function of other parameters such as position on the sky) and
 due to uncertainty in the Solar motion or the rotation curve used to calculate the eccentricity.\footnote{
At low eccentricity the difference  $E - E(L) \sim 0.5 \kappa^2 r_g^2 e^2$
where $r_g$ is the guiding radius, $e$ is the orbital eccentricity, $E$ is the orbital energy per unit mass,
$E(L)$ is the energy (per unit mass) of a circular orbit with angular momentum $L$,  
and $\kappa$ is the epicyclic frequency.  
For a power law rotation curve $v_c(r) \propto r^{-\alpha}$ 
the epicyclic freqency $\kappa = \sqrt{2(1-\alpha)}\Omega$ where the angular rotation rate $\Omega = v_c/r$.
Uncertainty in the slope of the rotation curve affects the estimate for $E(L)$ and
epicyclic frequency, $\kappa$, and so the computed values for the eccentricity.}

\begin{figure*}
\includegraphics[width=5.5in, trim= 0 0 0 0 ]{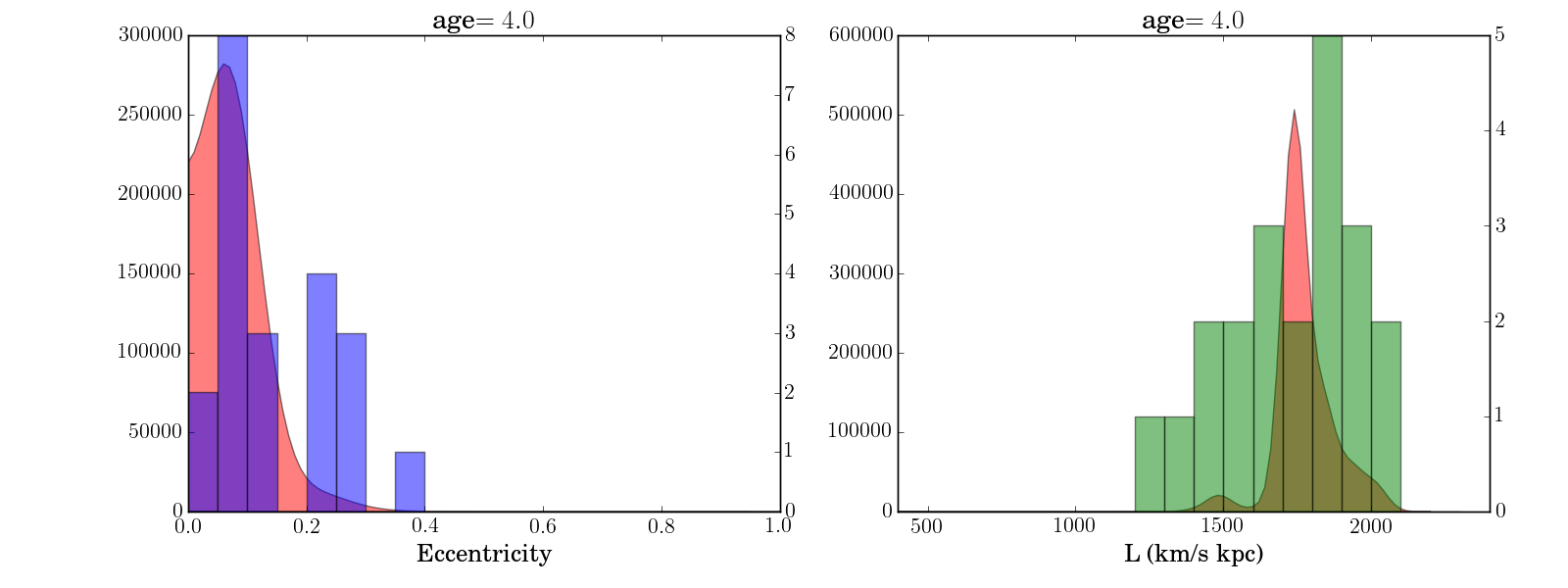}
\includegraphics[width=5.5in, trim= 0 0 0 0 ]{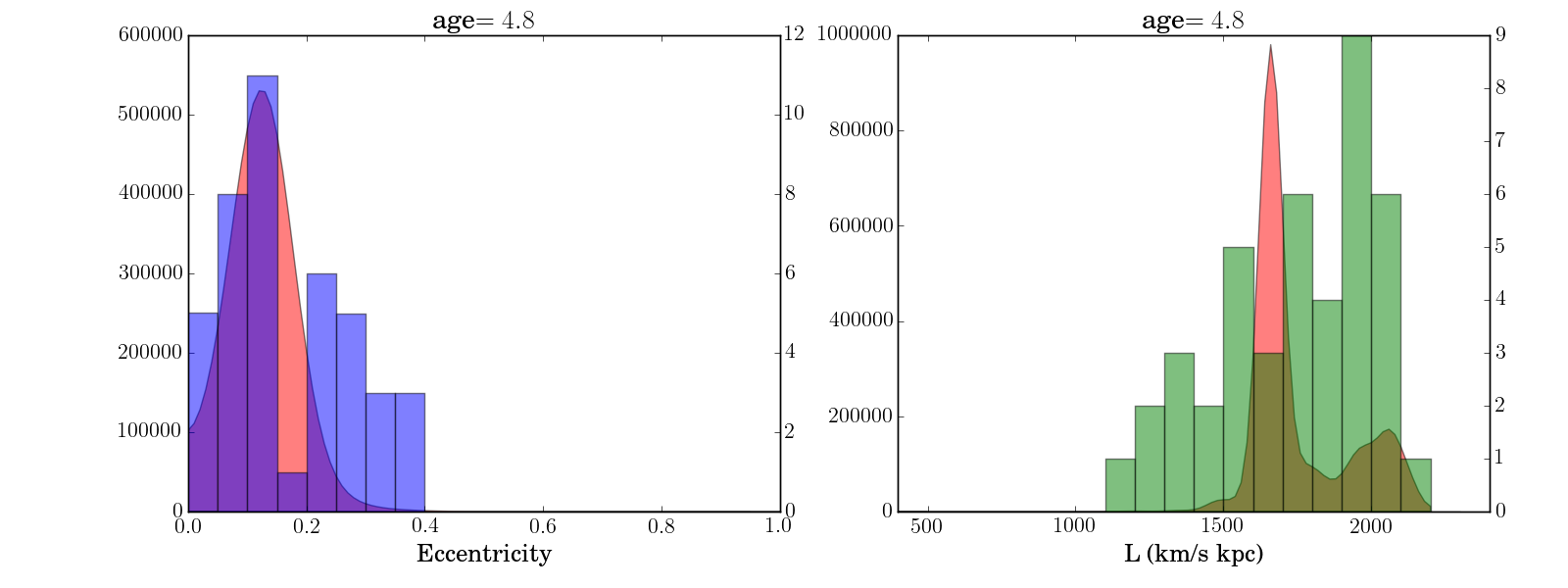}
\includegraphics[width=5.5in, trim= 0 0 0 0 ]{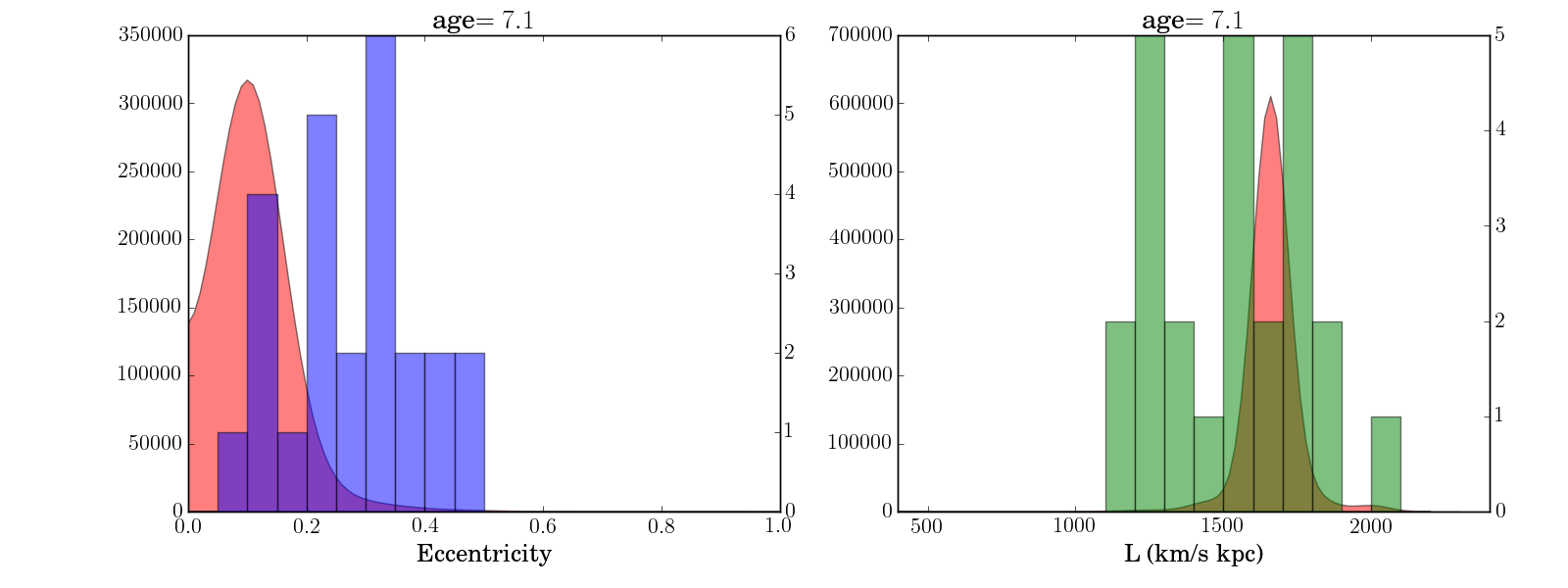}
\caption{Eccentricity and angular momentum distributions of the 6 groups.
The left panels show the eccentricity distributions and the right panels show the
angular momentum distributions.   Each row shows a different group, with group labelled by age.
Shown in blue with axis on the right hand side of the left panels are the observed eccentricity distribution
of each group, plotting numbers of stars in bins of width 0.05 in eccentricity.  
Shown in green, with axis on the right hand size of the right panels are
the angular momentum distributions of each group, showing the number of stars in bins of width 100 \kmsp.
The pink shaded regions, corresponding to axes on the left side of each panel, show the estimated
numbers stars in the
parent population distributions (corrected for selection and assumed to be evenly distributed in azimuthal angle, 
see Section 4) with bin widths of 0.01 in eccentricity or 20 \kmsp ~ in angular momentum.
\label{fig:hist}  } 
\end{figure*}

\setcounter{figure}{2}
\begin{figure*}
\includegraphics[width=5.5in, trim= 0 0 0 0 ]{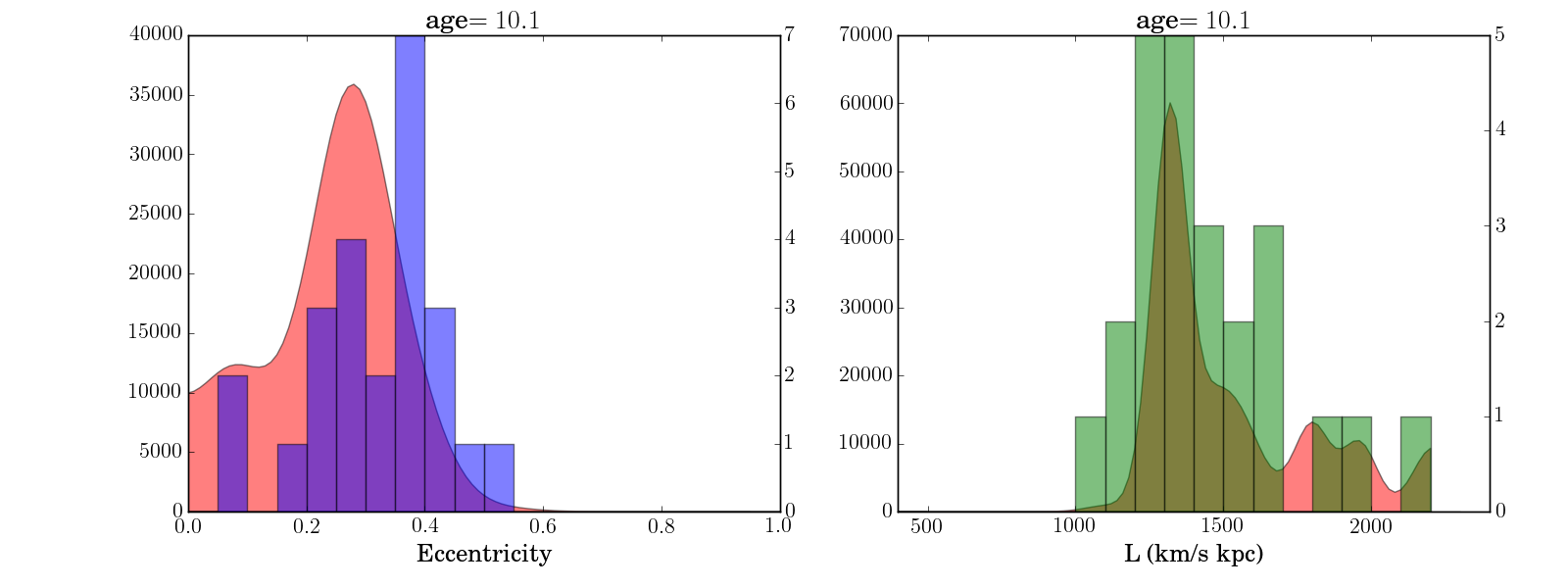}
\includegraphics[width=5.5in, trim= 0 0 0 0 ]{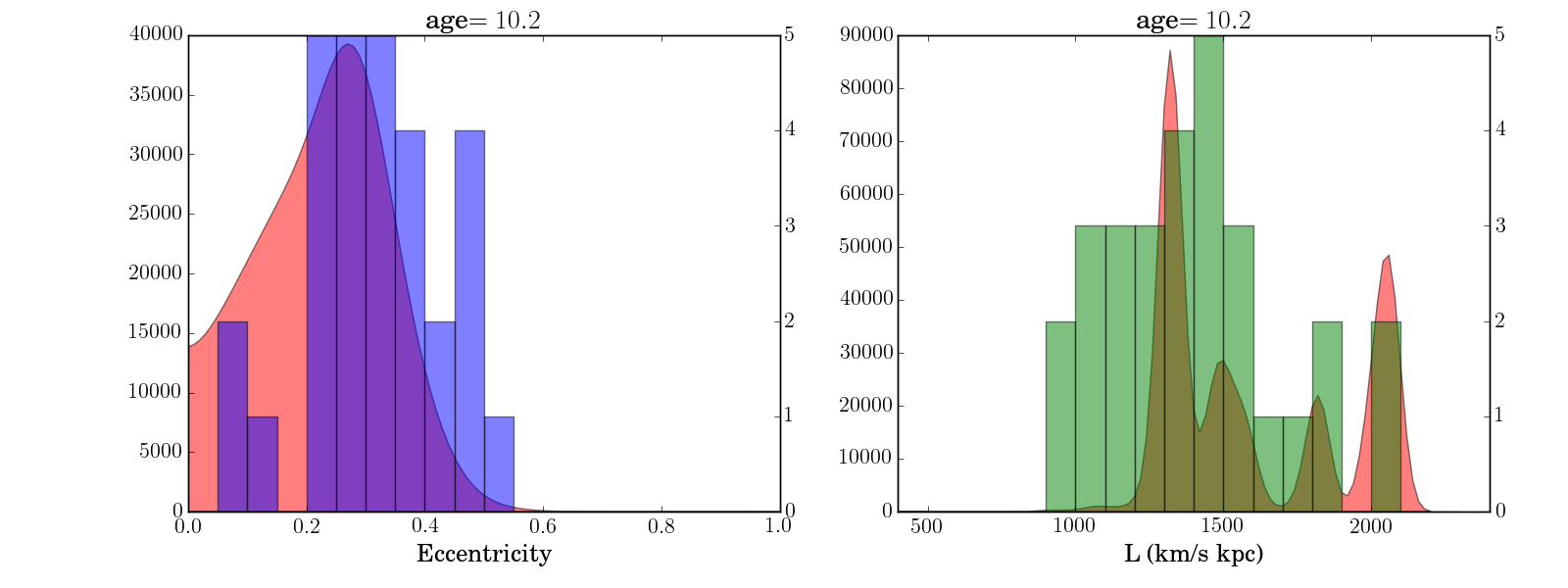}
\includegraphics[width=5.5in, trim= 0 0 0 0 ]{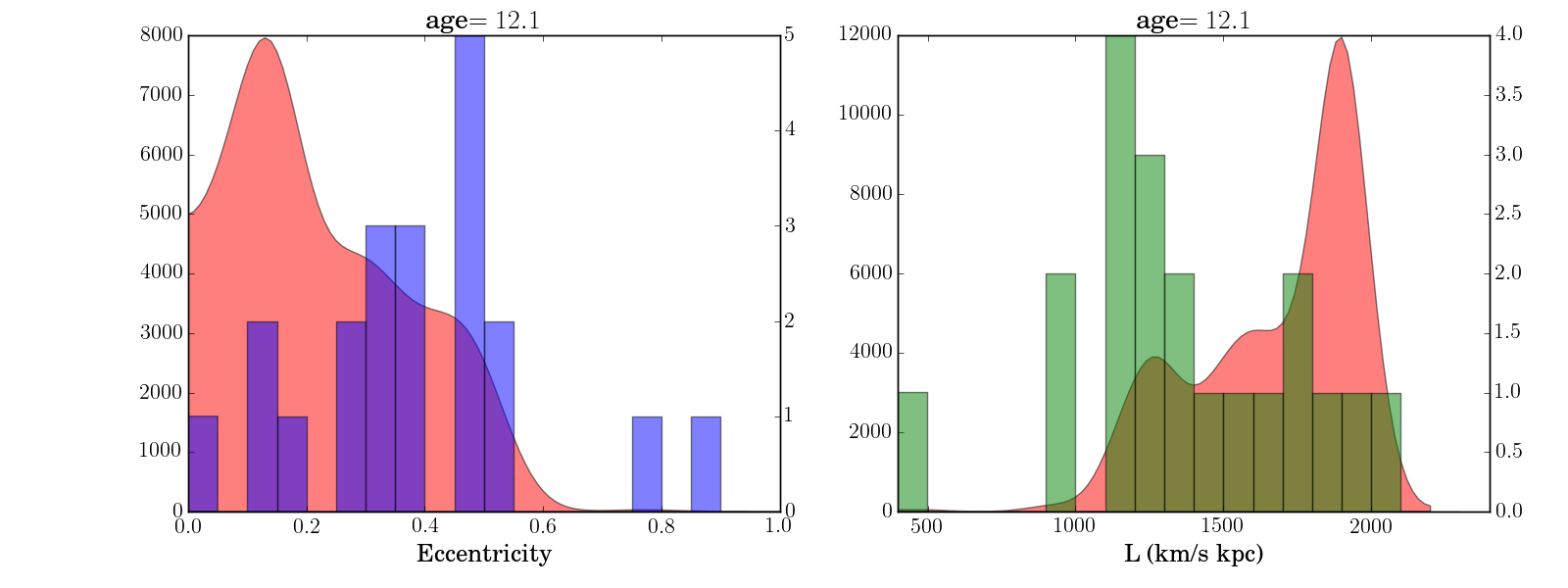}
\caption{continued}
\end{figure*}

\begin{figure*}
\includegraphics[width=5.0in, trim= 0 0 0 0 ]{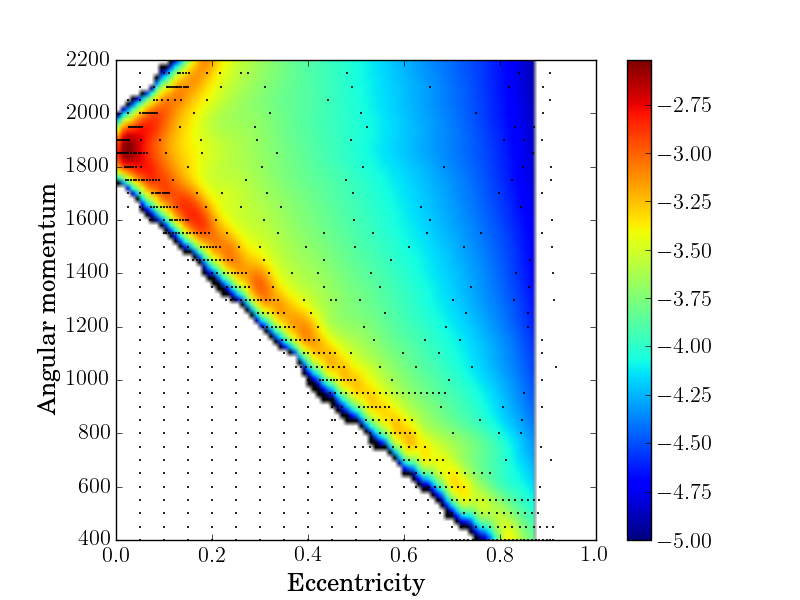}
\caption{As a function of orbital eccentricity and angular momentum we show $\log_{10} p_o(e,L)$ of the probability 
that a distribution of stars in such an orbit (but with randomly chosen angles) 
would be observed in the Solar neighborhood (100 pc from the Sun).
The probability is computed using planar orbits integrated in the Galactic potential model by 
\citet{allen91}.  Each orbit integrated is shown as a black dot. The angular momentum is in units of \kmsp.
\label{fig:lprb}  } 
\end{figure*}

\section{Probability of detecting a star  in the Solar neighborhood as a function of orbital eccentricity and angular momentum}

In this section we describe how to estimate the probability that a star with a given angular momentum and eccentricity
is found in a solar neighborhood sample if the parent population is randomly distributed in azimuthal and epicyclic angles.
We use angular momentum and eccentricity to describe each orbit.    We use eccentricity $e$ instead of energy as it is unitless, describes
the extent of radial excursion in the orbit and so gives an intuitive description for the orbit shape, 
and it does not depend on a potential energy offset.

To be consistent with the angular momentum and eccentricities 
computed by \citet{bensby14} and \citet{mitschang14} we use the same Galactic potential model as they did to compute our probabilities.
Using the gravitational potential  for the Galactic model
by \citet{allen91}\footnote{This model assumes a Galactocentric distance for the Sun and rotation velocity of a circular orbit
 at that radius of $R_\odot = 8.5$ kpc and $V_{LSR,\odot}  = 220$ km~s$^{-1}$.  With these values the angular momentum 
 of the local standard of rest is $L_{LSR} = 1870$ \kmsp.}
 we integrate planar orbits with different initial angular momentum and different initial radii.
For each orbit we record 
the eccentricity defined as $e=(R_a - R_p)/(R_a + R_p)$ (following \citealt{bedin06}) where $R_a, R_p$ are radii of
galactic apocenter and pericenter, respectively.

For each $e,L$, we computed
a few thousand positions in a full orbit (using a finite size timestep to compute a full orbital period). We then 
randomly chose a few thousand azimuthal angles (corresponding to randomly chosen initial orientations) 
giving a total of approximately 10 million 
points in the galaxy plane to compute each probability.   At each timestep and for each angle we computed
the position of the star and the fraction that fell within a solar neighborhood area,  within 100 pc of the Sun,
gave the probability. 
For randomly distributed initial azimuthal angle and initial position in the orbit, and using a single orbital period, 
we measure the fraction of stars
in an orbit, as a function of  angular momentum and eccentricity, that are located within 100 pc
of the Sun. 
In other words, we assume there is a distribution of orbits with this angular momentum and eccentricity that is 
randomly distributed in  azimuthal angle, and 
using this distribution, we
compute the probability, $p_o(e,L)$, that a star would be observed in the Solar neighborhood at any particular time.
For a range of orbital eccentricities, $e$, and angular momenta, $L$, we compute $p_o(e,L)$
and display it in Figure \ref{fig:lprb}.  The color bar shows the $\log_{10}$ of the probability.
The black dots show the orbits that we integrated  and that were
used to make the color contours.  The angular momentum is in units of \kmsp.
Wiggles in Figure \ref{fig:lprb} are artifacts due to the sampling of the orbits integrated.

\begin{figure*}
\includegraphics[width=3.2in, trim= 0 0 0.2in 0 ]{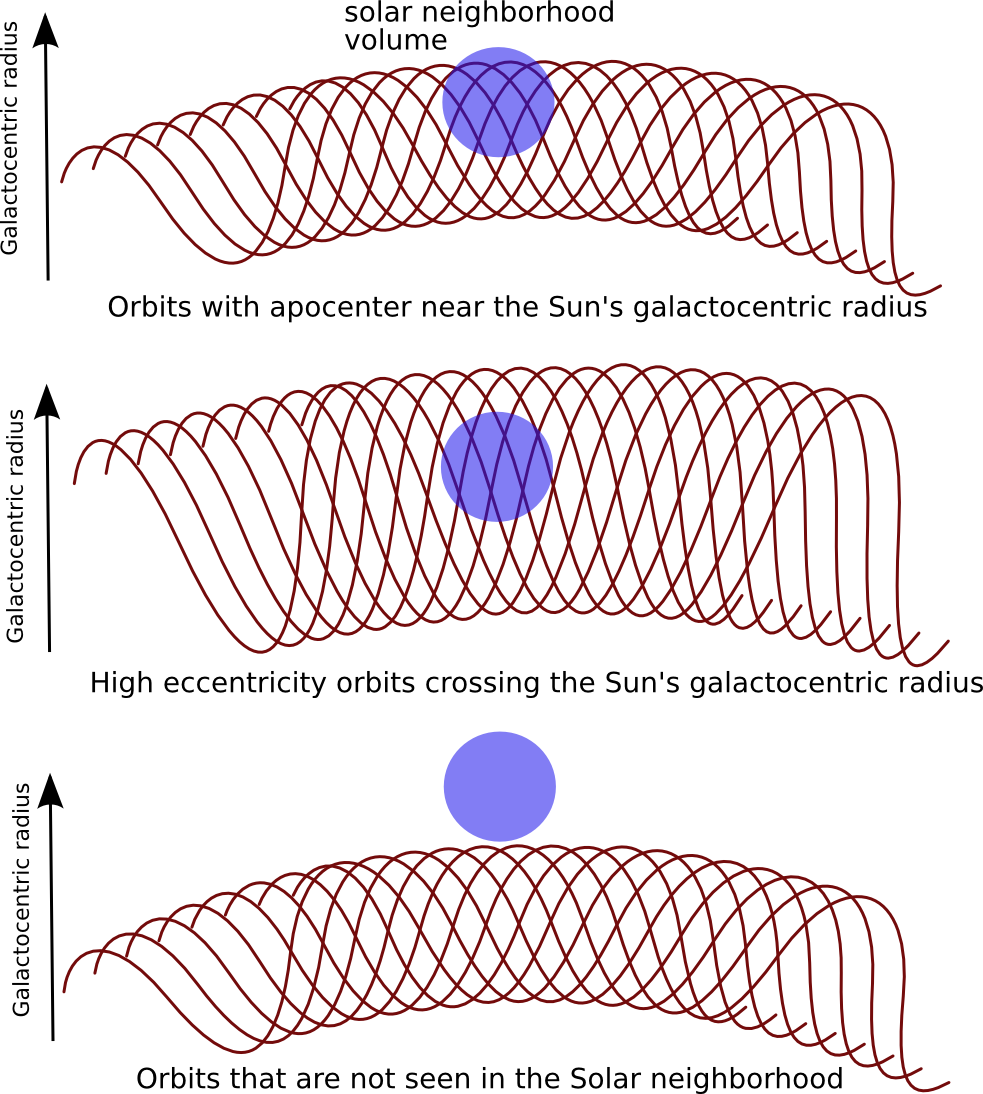}
\caption{Orbits with apocenter near Sun's galactocentric radius (as shown on the top) are more likely to be
seen in a solar neighborhood sample, than a high eccentricity orbit (as shown in the middle).
The bottom illustrates orbits that never are found in the Solar neighborhood, corresponding
to a region in $e,L$ space that we call  the forbidden region.
\label{fig:orbit}
}
\end{figure*}

At low eccentricity and angular momentum $L$ above or below that of the local standard of rest, 
the probability $p_o$, for the orbit is zero as the orbit never crosses  the Sun's galactocentric radius.
A star in such an orbit is never near the Sun.   
The white region on the lower left and upper left in Figure \ref{fig:lprb} is this forbidden region. 
Large eccentricity orbits that do cross the Sun's galactocentric radius (on the right in Figure \ref{fig:lprb})
are less probable than lower eccentricity ones as stars spend much of the time at larger or smaller galactocentric radius
than that of the Sun.  
For a given angular momentum, the probability is highest at an eccentricity that just barely allows 
the orbit to cross into the Solar neighborhood.  We attribute the increase in probability near the forbidden
region boundary to the large fraction of the orbital period spent near a particular radius when at apocenter or pericenter.
This effect has previously been described as a bias due to {\it crossing times} in the Solar neighborhood \citep{mayor77}.
The effect is illustrated in Figure \ref{fig:orbit} showing epicyclic oscillations for three different groups of orbits, one with apocenter
near the Sun's galactocentric radius that is likely to be seen in the Solar neighborhood, high eccentricity orbits
that have a lower probability and orbits within the forbidden region that cannot be found in Solar neighborhood.

We can account for this probability increase near apocenter
using an epicyclic approximation for radial orbital variations.
For low eccentricity stars the radius 
$r(t) \approx r_g (1 + e \cos (\kappa t + \phi_0))$ where $r_g$ is the guiding radius,
$e$ is the eccentricity,  $\kappa$ is the epyclic frequency, $\phi_0$ an initial phase and the apocentre radius
$R_a = r_g(1+e)$.
Near apocenter and using a small angle approximation,
$R_a - r(t) \propto (t - t_{apo})^2$ where $t_{apo}$ is a time when the orbit is at apocenter.  
This gives a dependence of the fraction of the orbital period, $f$,
spent within a narrow annulus of width $dr$ from apocenter,  $f \propto \sqrt{dr}$.   In contrast when
the orbit is near the guiding radius and using a small angle approximation,
 $r - r_g \propto (t - t_g)$ (with $t_g$ a time the orbit crosses the guiding radius) 
giving a dependence of the fraction 
of the orbital period spent within $dr$ of $r_g$ to be $f \propto dr$.   For a small range of radius $dr$,
the fraction of the orbital period spent near apocenter is larger than that spent near the middle of the orbit 
at the guiding radius.  The trend is still present at moderate eccentricity where the epicyclic approximation
is less accurate.

\subsection{Probabilities of individual abundance group stars} 
\label{sec:prob_ind}

\begin{figure*}
\includegraphics[width=3.2in, trim= 0 0 0.2in 0 ]{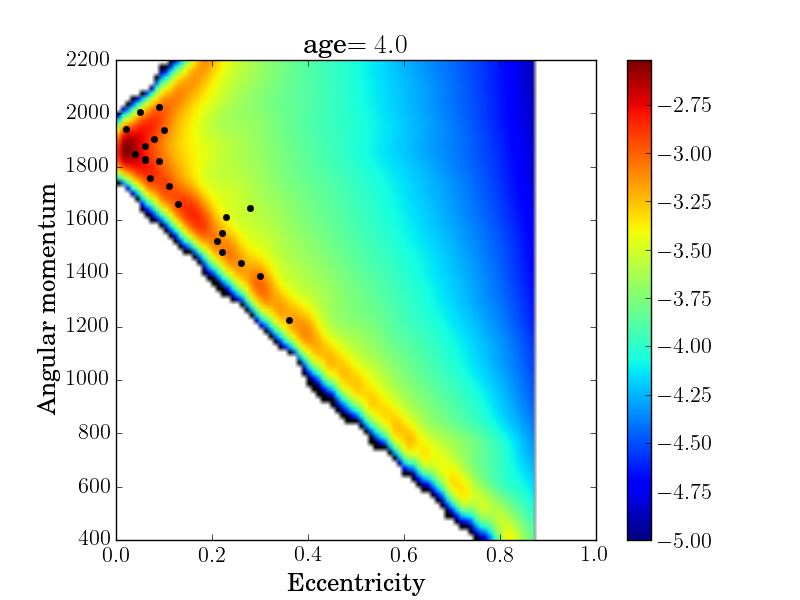}
\includegraphics[width=3.2in, trim= 0 0 0.2in 0 ]{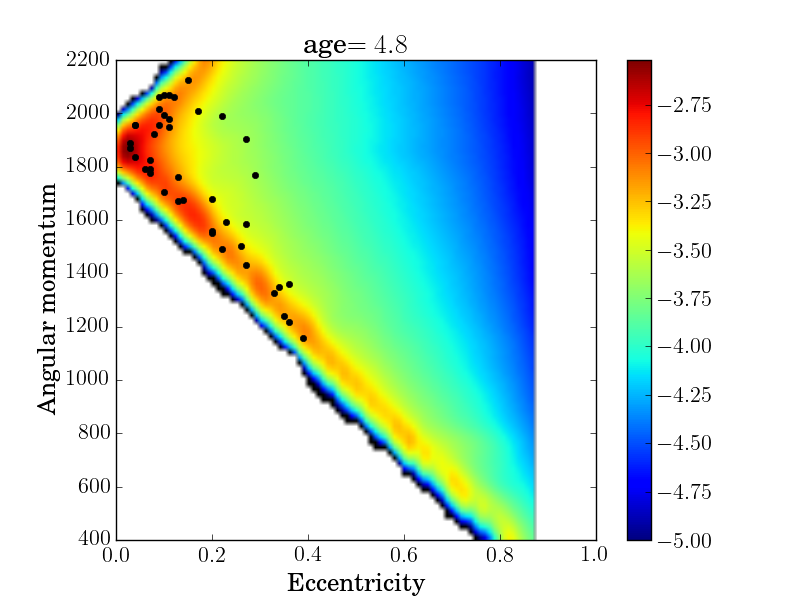}
\includegraphics[width=3.2in, trim= 0 0 0.2in 0 ]{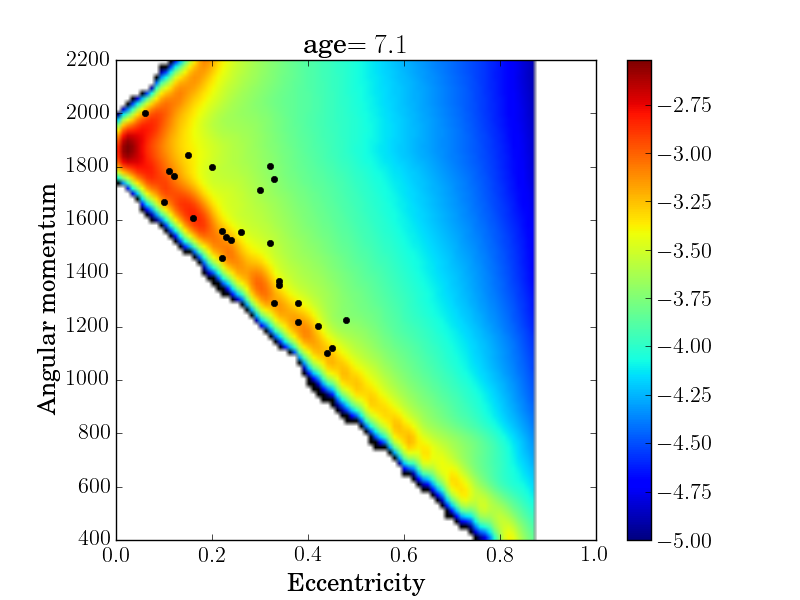}
\includegraphics[width=3.2in, trim= 0 0 0.2in 0 ]{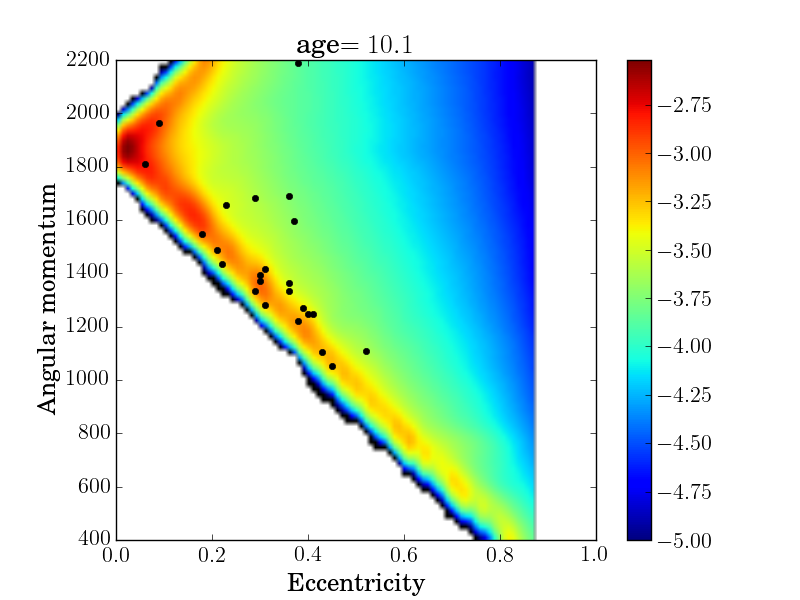}
\includegraphics[width=3.2in, trim= 0 0 0.2in 0 ]{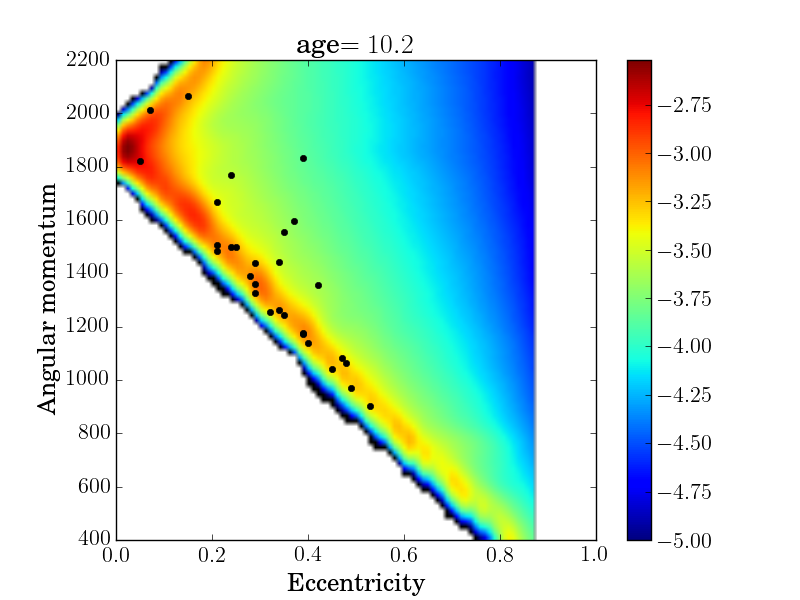}
\includegraphics[width=3.2in, trim= 0 0 0.2in 0 ]{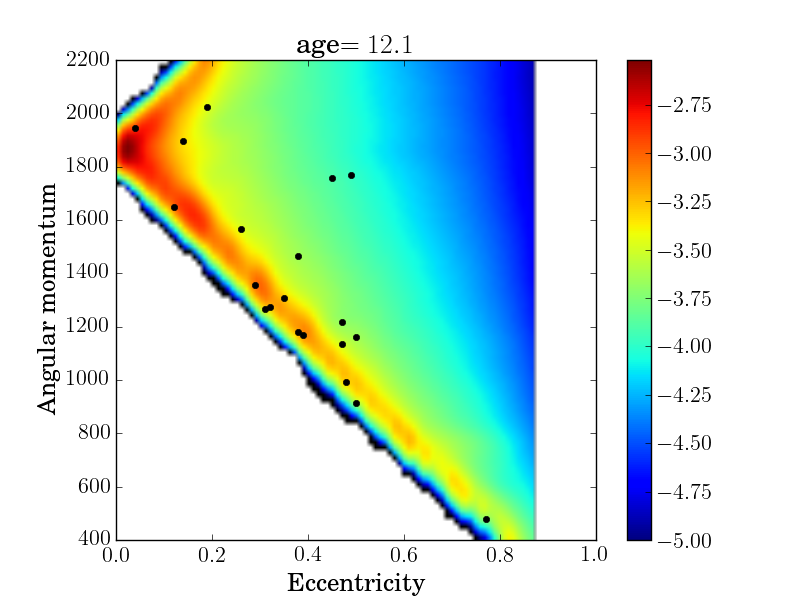}
\caption{Eccentricity and angular momentum of stars in an abundance group plotted as points on
top of the probability that such an orbit is detected in the Solar neighborhood.
Each panel shows a different abundance group and the groups are labelled by their ages.
The color bar shows $\log_{10}p_o$ of the probability computed for each orbit.
The stars in each abundance group often lie in regions of high probability, near the boundary of the forbidden region.
\label{fig:probc}  } 
\end{figure*}

We now consider the probabilities that stars in the abundance groups are seen in the Solar neighborhood.
In Figure \ref{fig:probc} we show the eccentricity and angular momentum of stars in these 
abundance groups on top of the probability, $p_o$, that such a star is found in the Solar
neighborhood.  Each panel shows a different abundance group and the groups are labelled by their ages.
The probability is displayed as in Figure \ref{fig:lprb}.
Figure \ref{fig:probc} illustrates that  stars tend to be found near the forbidden region in  $e, L$ space,
as expected from the location of the high values in probability distribution $p_o(e,L)$.

For each star  with eccentricity $e_i$ and angular momentum $L_i$ we can
use the probability $p_o(e_i,L_i)$ to estimate the size of the parent grouping.
The parent population has at least
\begin{equation}
N_1 = \sum_i \frac{1}{ p_o(e_i, L_i)}  \label{eqn:N1}
\end{equation}
stars in it.
 If we underestimate the probability $p_o(e_i,L_i)$ then we will overestimate the number of
 stars in the parent population.  
 To ensure that observational errors in $e_i$ and $L_i$ for individual stars do not give spurious high
 numbers near the forbidden region, we take $p_o$ to be the maximum value within  $e_i  \pm \Delta e$
 and $L_i \pm \Delta L$ with $\Delta e=0.01$ and $\Delta L = 13$ \kmsp, the size of  the errors estimated for these 
 quantities (see end of section 2).
For each abundance group, we have summed the inverse of the probabilities and list the total
number of estimated parent stars in Table \ref{tab:parents}.

\section{Parent population distributions}

\begin{figure*}
\includegraphics[width=3.2in, trim= 0 0.25in 1.4in 0 ]{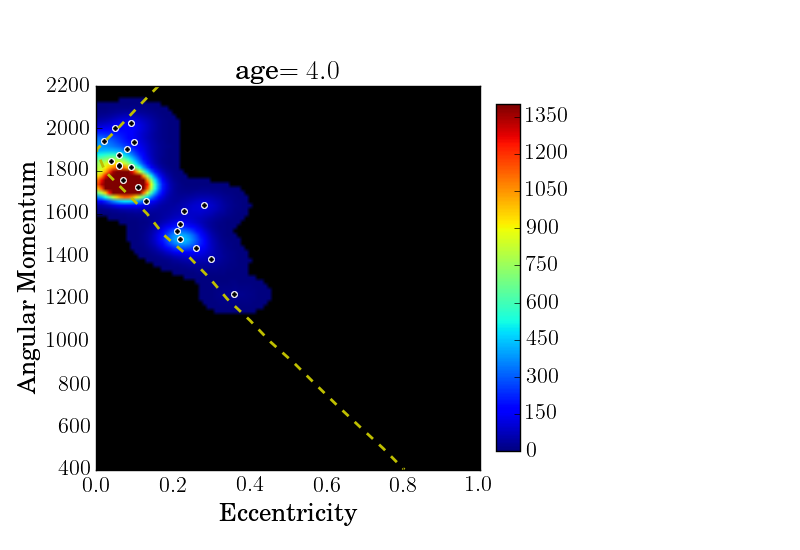}
\includegraphics[width=3.2in, trim= 0 0.25in 1.4in 0 ]{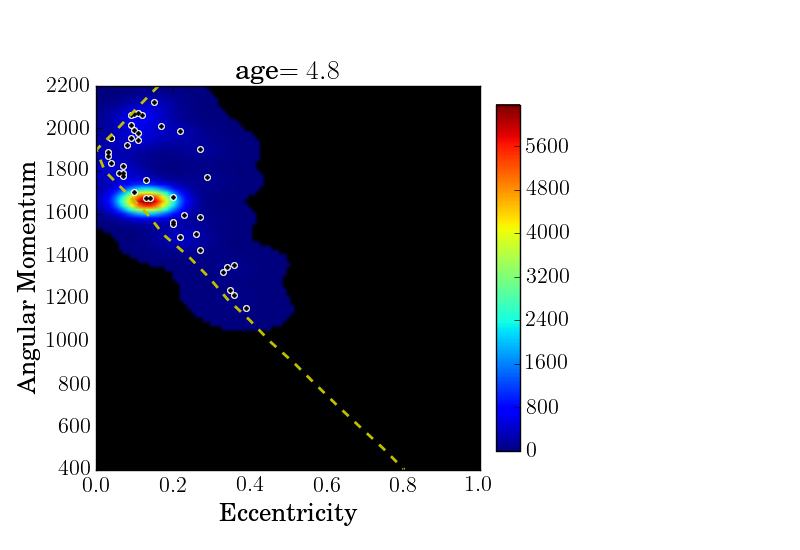}
\includegraphics[width=3.2in, trim= 0 0.25in 1.4in 0 ]{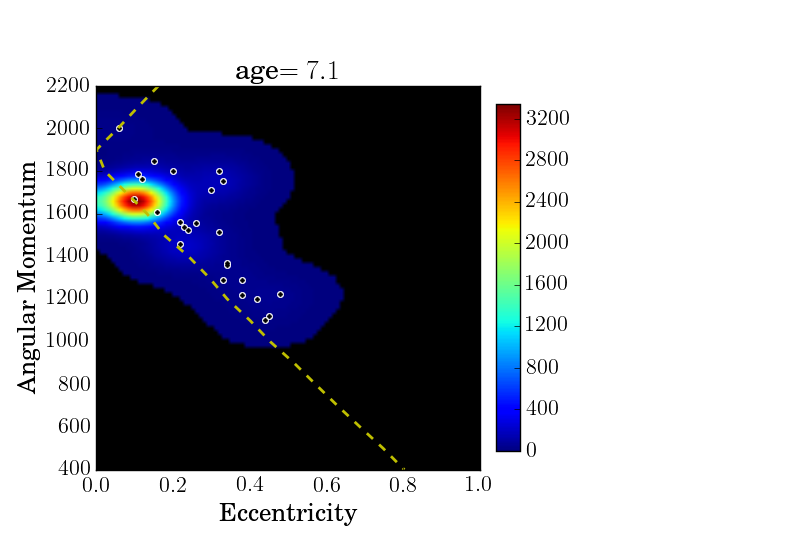}
\includegraphics[width=3.2in, trim= 0 0.25in 1.4in 0 ]{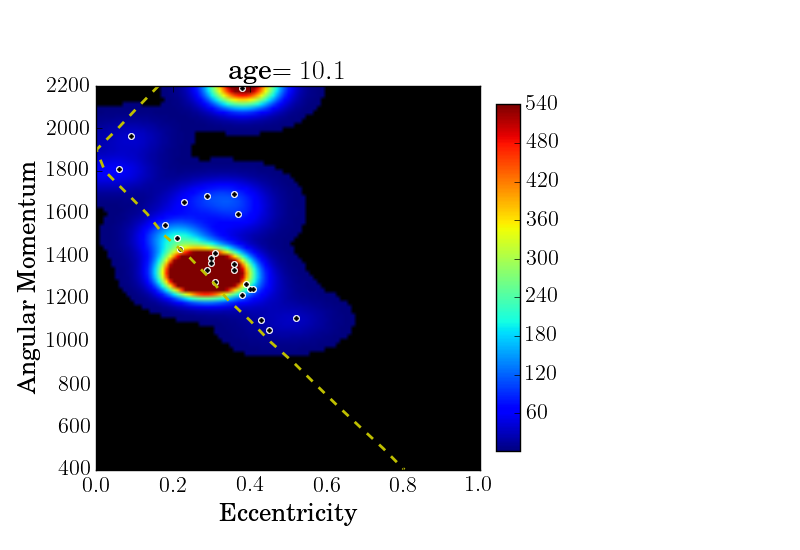}
\includegraphics[width=3.2in, trim= 0 0.25in 1.4in 0 ]{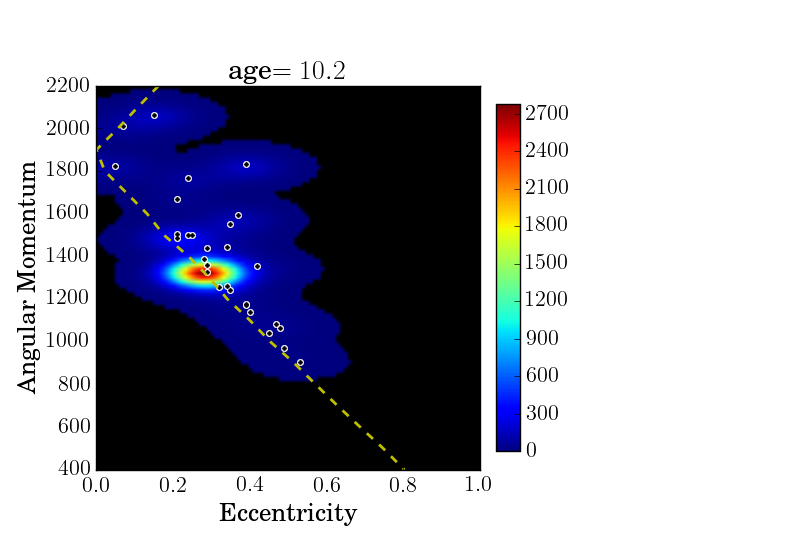}
\includegraphics[width=3.2in, trim= 0 0.25in 1.4in 0 ]{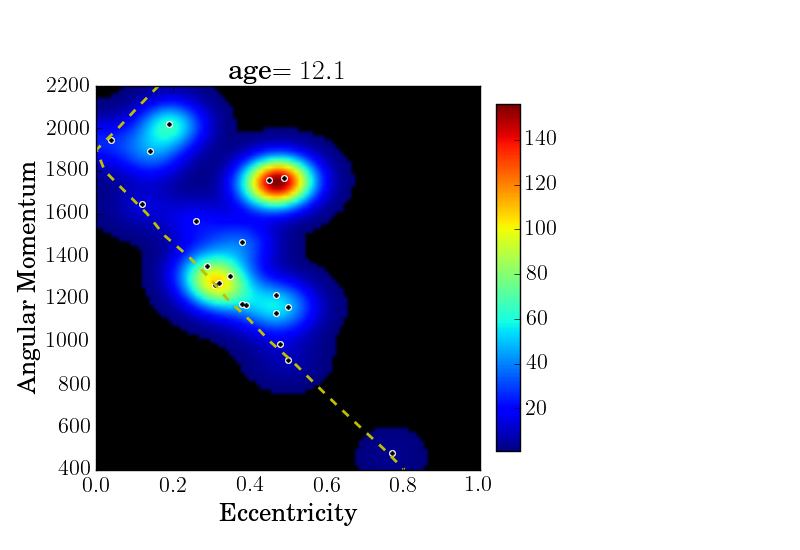}
\caption{Distribution of parent populations of abundance groups, taking into account the orbital probability and using the eccentricities
and angular momenta of the stars in the group.  Each group can be identified by its age on the top of the plot.
Eccentricity and angular momentum of stars in each abundance group are plotted as points on
top of number density of the estimated parent distribution.  
Each panel shows a different abundance grouping with group properties listed in Table \ref{tab:groupings}.
The color bar shows the estimated number of stars per eccentricity and angular momentum bin with
bin width and height $de = 0.01$ and $dL = 20$  \kmsp.  The dashed yellow lines border the forbidden region.
\label{fig:probi}  } 
\end{figure*}

Using the probabilities estimated for each star in an abundance grouping we now estimate the number density
of stars in the grouping as a function of eccentricity and angular momentum $e, L$.
Each star with $e_i, L_i$ contributes a total parent population of $p_o(e_i,L_i)^{-1}$ at $e_i, L_i$.
We smooth this distribution to estimate the number density of stars in the parent population as a function 
of $e,L$.  For each cluster, the resulting distributions  in $e$ and $L$ are shown in Figure \ref{fig:probi}.
The color bars show the number of stars per eccentricity and angular momentum bin with
bin size $de = 0.01$ and $dL = 20$ \kmsp.  
The distributions have been smoothed by 4 or 5 eccentricity bins and 2 or 3 angular momentum bins
with the tighter distributions (for the younger groups) smoothed by fewer bin widths.
 
\begin{figure*}
\includegraphics[width=3.2in, trim= 0 0 0 0 ]{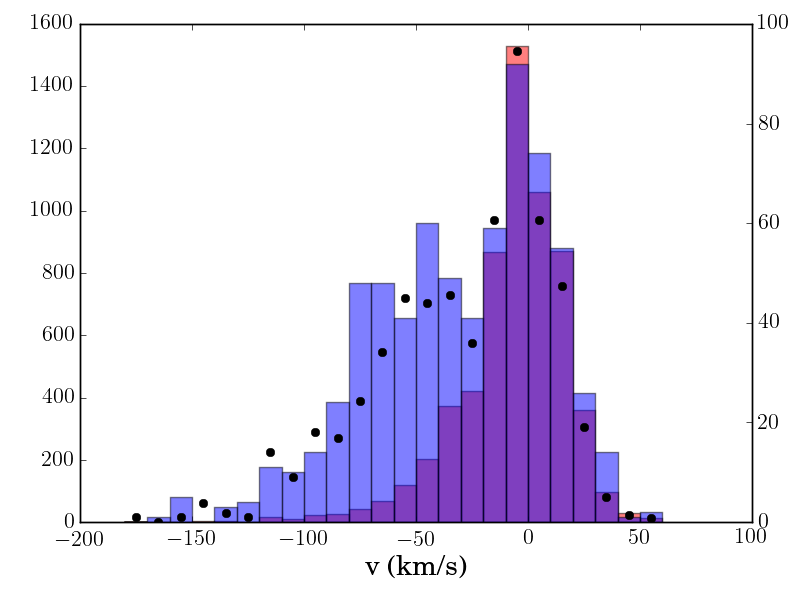}
\caption{The angular momentum distribution of the \citet{bensby14} sample is shown in blue (with axis on the right)
and that of the Geneva Copenhagen Survey (GCS) in red (with axis on the left) with overlap regions 
shown in purple.  The distributions are shown as a function of the $V$ tangential velocity component and using 10 \kmsp~ bins.
The vertical axes show the number of stars in these bins.  
The GCS histogram when multiplied by the function in equation \ref{eqn:fL} is shown with black dots (with axis on the right).
The black dots match the \citet{bensby14} histogram giving us an estimate of the selection bias compared to the GCS.
\label{fig:sel}  } 
\end{figure*}

 \citet{bensby14} selected stars for spectroscopic study so that the sample contained extremes
 of both thin and thick disk.
 Therefore, many thin disk stars were necessarily neglected from
 the \citet{bensby14} sample.
 As a result, the sample contains a bias against high angular momentum stars.
 In Figure \ref{fig:sel}, 
 we compare the angular momentum distribution of the \citet{bensby14} sample
  to that of the GCS stars \citep{holmberg09},
 but restricted to stars within 80 pc.
  In this figure the blue histogram shows the \citet{bensby14} sample
 and the red histogram the GCS stars with overlaps displayed as purple.
 
\begin{figure*}
\includegraphics[width=3.2in, trim= 0 0 0 0 ]{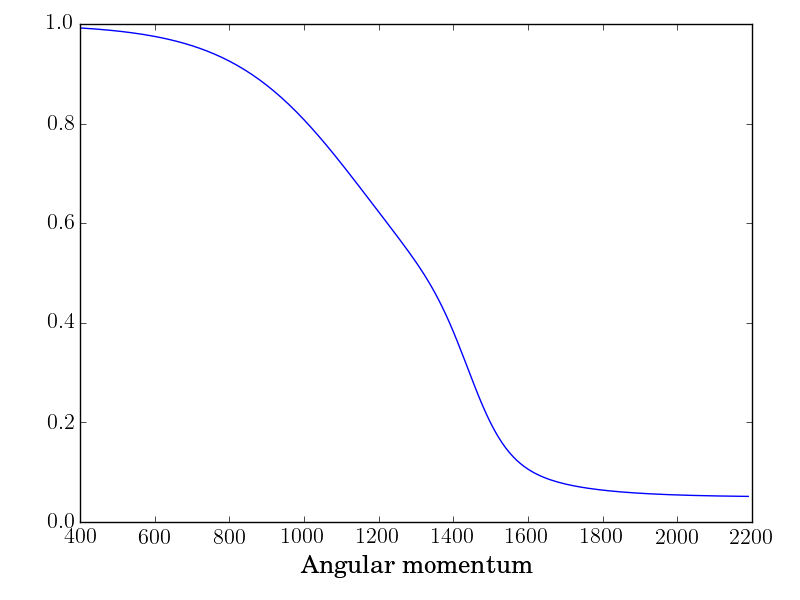}
\caption{The selection function $f(L)$ (equation \ref{eqn:fL}) is shown here as a function of angular momentum
in \kmsp.  The \citet{bensby14} sample contains proportionally more low angular momentum stars than the Geneva
Copenhagen Survey.
\label{fig:sel_L}  } 
\end{figure*}

We constructed a selection function, $f(L)$,  
choosing  two $\tanh$ functions as they go smoothly between
one constant to another constant value and this allows us to model the
 the two humps in the \citet{bensby14} sample evident in the angular momentum distribution shown in Figure \ref{fig:sel}.
 The function we chose is described with a few parameters, is smooth, 
 is never extremely small and does not cross zero (this is important as we need to divide by it).
The black dots in Figure \ref{fig:sel} show the GCS histogram multiplied by $f(L)$ with 
 \begin{equation}
 f(L) = a_0 -  \frac{a_1}{2}\tanh \left( \frac{L - L_1}{s_1 L_{LSR}} \right)
 - \frac{a_2}{2}\tanh \left(\frac{L - L_2}{s_2 L_{LSR}}\right) \label{eqn:fL}
 \end{equation}
and coefficients $a_0 = 0.53, a_1 = 0.65, a_2=0.30, s_1 = 0.18, s_2=0.05$ and $L_1=1148, L_2 = 1445$ \kmsp~
and with  $L_{SLR} = 1870$ \kmsp.
The function itself is plotted in Figure \ref{fig:sel_L}.   
 We did not automatically fit the coefficients, but did adjust the coefficients
so that the two histograms lay on top of one another.
Because we divide by this function, if it is an underestimate for the selection of the Bensby et al. sample, 
then we will overestimate the size of parent populations.

 \renewcommand{\arraystretch}{1.2}
 The function $f(L)$ is an estimate for the fraction of stars selected by \citet{bensby14}
 compared to that in the Geneva Copenhagen Survey.
 At $L$ near that of the local standard of rest, 
$f(L) \sim 0.05 $ implying that for every high $L$ star in the \citet{bensby14} sample
 there are 20 stars with similar angular momentum in the GCS.  
 At high angular momentum,  the \citet{bensby14} sample also contains a higher proportion of high eccentricity
 stars than the GCS.   To ensure that we don't overestimate the number of high angular momentum high eccentricity 
 stars the parent populations we cut the selection function with 
 \begin{eqnarray}
 f(L,e) &=&   1 {~ \rm for~}  L>1600 ~{\rm km~s^{-1}~kpc} {\rm ~and~}  e >0.15 \nonumber \\
 &=& f(L) {\rm ~~~~ otherwise} 
  \end{eqnarray}
 
\begin{figure*}
\includegraphics[width=3.15in, trim= 0.2in 0.25in 1.4in 0 ]{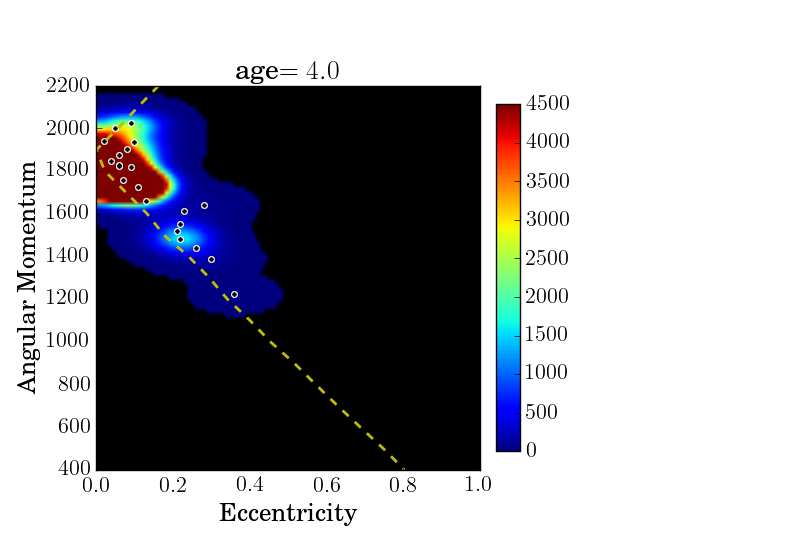}
\includegraphics[width=3.75in, trim= 0.2in 0.25in 0 0 ]{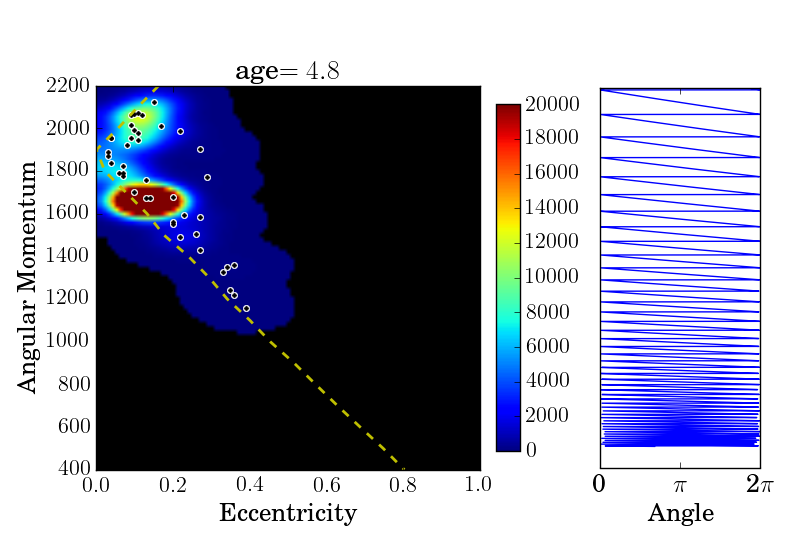}
\includegraphics[width=3.15in, trim= 0.2in 0.25in 1.4in 0 ]{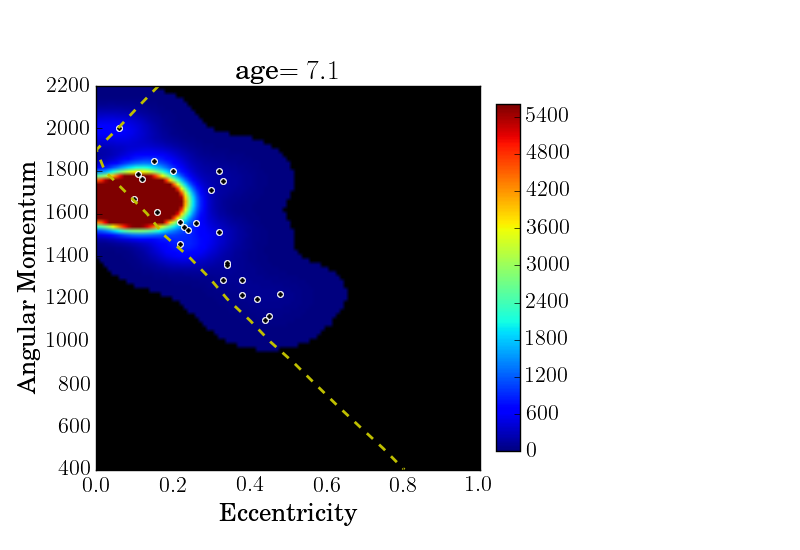}
\includegraphics[width=3.75in, trim= 0.2in 0.25in 0 0 ]{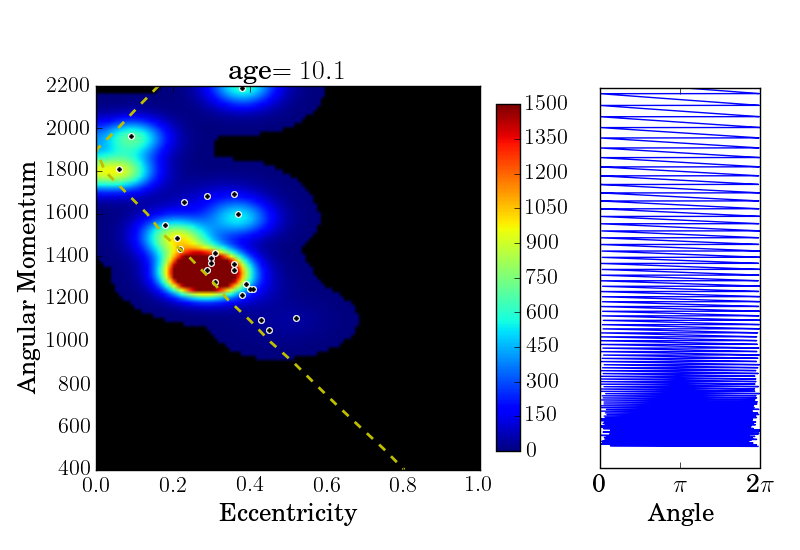}
\includegraphics[width=3.15in, trim= 0.2in 0.25in 1.4in 0 ]{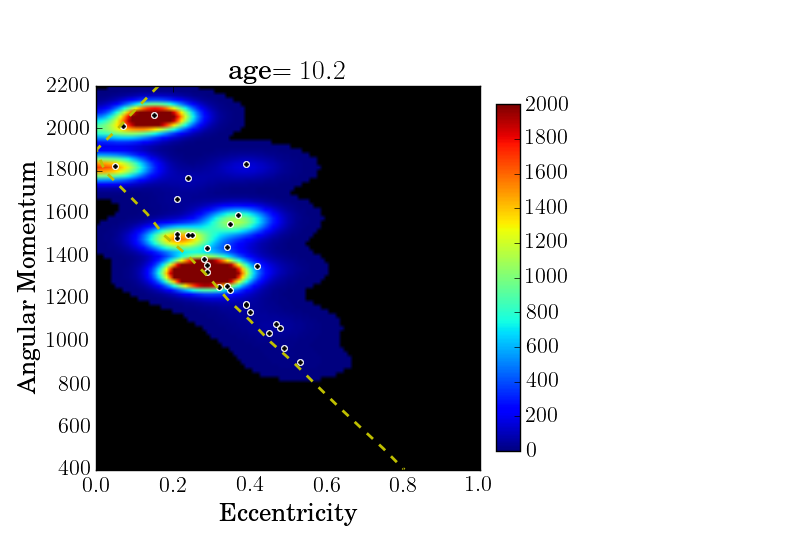}
\includegraphics[width=3.75in, trim= 0.2in 0.25in 0 0 ]{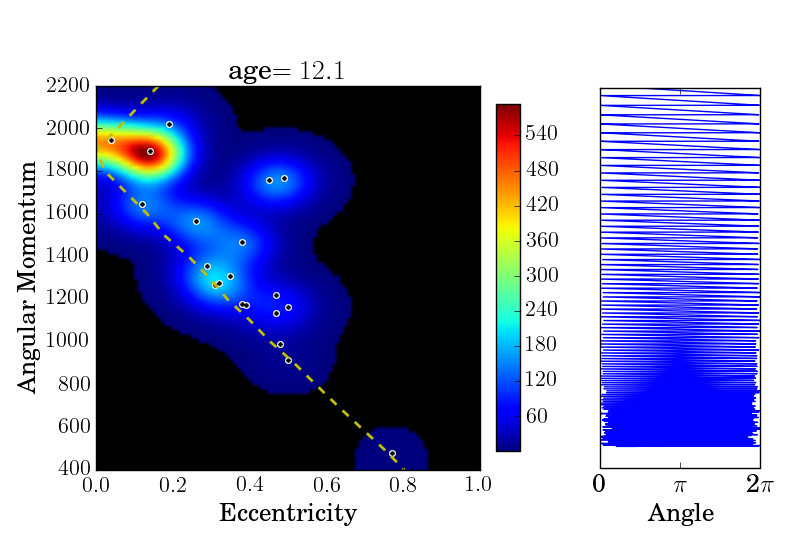}
\caption{Distribution of parent populations of abundance groups.
Similar to figure \ref{fig:probi} except a bias against high angular momentum stars has been
removed.  Each panel shows a different abundance grouping with group properties listed in Table \ref{tab:groupings}.
Eccentricity and angular momentum of stars in each abundance group are plotted as points on
top of number density of the estimated parent distribution.   The dashed yellow lines border the forbidden region.
On the right for three of the groups, we show an additional panel plotting
the azimuthal angle $\theta(L)$ for a group originally at the same angle but after a time
 equal to the age of the group.  
The $y$-axis on these rightmost panels is angular momentum using the same scale as for the abundance distribution 
and the $x$-axis is the azimuthal angle $\theta$.  Only when $\theta(L)$ is slowly varying should we have overestimated the parent
population by assuming an axisymmetric distribution.   We have likely only overestimated the parent population size for the two youngest
groups and only at angular momentum near 1800 \kmsp.
\label{fig:probj}  } 
\end{figure*}

 We correct the probability for each star in each abundance grouping
 with this selection function giving a total number for the parent population
 \begin{equation}
 N_2 = \sum_i \frac{ 1} { p_o(e_i, L_i) f(L_i,e_i) } \label{eqn:N2}
 \end{equation}
 again taking the minimum value for $p_o$ within $e_i \pm \Delta e$ and $L_i \pm \Delta L$.
The distribution of the parent populations for each abundance grouping, also taking
into account the selection function, are shown in Figure \ref{fig:probj} and the numbers $N_2$
listed in Table \ref{tab:parents}.
As expected the total number of stars estimated for each group is larger than that estimated
previously without using the selection function. 
We integrate the parent distributions plotted in Figure \ref{fig:probj} in eccentricity to estimate angular momentum
distributions.  Likewise integrating in angular momentum we can estimate eccentricity 
distributions.  The estimated parent eccentricity distributions and the parent angular momentum distributions
are shown as pink solid regions in Figure \ref{fig:hist} where numbers of stars in eccentricity bins of size 0.01
are plotted in the left panels and numbers of stars in angular momentum bins of size 20 \kmsp ~ are plotted in
the right panels.

\subsection{Discussion on azimuthal structures and phase wrapping}

To estimate the probability $p_o$ we have assumed that the parent population is evenly distributed azimuthally in the galaxy. 
However,
an originally cold disrupted cluster may not have time to become evenly distributed in azimuthal angle 
(for illustration see Figure 2 by \citealt{simon09}).
This would lead to a bias -- a survey of the Solar neighborhood would not see every group that is present
at the Sun's galactocentric radius.  Furthermore,
the parent population size of a group
detected in the Solar neighborhood would be overestimated by wrongly assuming that the group extended to all azimuthal angles.

Figure \ref{fig:hist} shows the eccentricity and angular momentum distributions of each group compared to
the distributions estimated in the parent populations.  This figure shows that the stars in a single group do not have the same
angular momentum.  The rotation period of a star
in the galaxy can be estimated from a star's angular momentum.  A spread in angular momentum
in the group implies differential rotation between the higher and lower angular momentum members of the group.
We consider how long it would take a disrupted cluster, with stars originally  at the same azimuthal angle
but with different angular momenta to shear out so that stars are located at every azimuthal angle in the Galaxy.

Because the angular rotation rate  $\Omega \sim v_c^2/L$ is approximately inversely proportional to
the angular momentum, (with $v_c$ the circular velocity and for an approximately flat rotation curve), 
the time  it takes an initially compact cluster with a spread in angular momentum values $dL$ to shear by $2 \pi$
in azimuthal angle is 
\begin{equation}
  \Delta t  \approx P \frac{L}{dL}   \end{equation}
with $P$ the mean rotation period  of the cluster.
The rotation period at the Solar neighborhood is $\sim 0.24$ Gyr.
In 4 Gyr there have been approximately 16 rotation periods giving $dL/L \sim 0.06$ for a group
that has sheared by  $2 \pi$ and is now distributed at all azimuthal angles.
Using  a solar value of $L_{LSR} = 1870$ for the cluster mean, we estimate $dL = 120$ \kmsp ~
is required for the cluster to shear to $2 \pi $ at $\Delta t = 4$ Gyr.
A parent population with a distribution with dispersion $dL \lesssim 120$ \kmsp ~ and age of 4 Gyr would not
be evenly distributed in azimuthal angle.    However, older populations with larger angular momentum
dispersions would be evenly distributed in the Galaxy. 

We use the rotation curve by \citet{allen91}\footnote{The second term of equation 5 by \citet{allen91} should have the opposite sign.} 
to compute  the azimuthal angle 
\begin{equation} 
\theta(L) = \Omega(L) \Delta t \label{eqn:theta}
\end{equation}
 as a function of angular momentum for a population that is initially 
at the same azimuthal angle at birth.  Here $\Delta t$ is the age of the group and $\theta$
is computed modulo $2 \pi$.  After $\Delta t$, the more rapidly rotating  stars (at lower angular momentum) will have increased
in $\theta$ more than those rotating slower (at higher angular momentum). 
For three of the groups in Figure \ref{fig:probj} we show the 
angle $\theta(L)$ as a side panel.    For the old groups, $\theta$ increases rapidly over a small
change in angular momentum.  As the angular momentum distributions for the old groups
are large, they are likely to be well distributed in the Galaxy.
In contrast, the younger groups contain peaks in the estimated parent populations that
are narrow in angular momentum width, and $\theta$ varies relatively slowly across that width.
In the peaks of the youngest two groups, we may have overestimated the parent populations
by a factor of a few if they are not evenly distributed in the Galaxy.
While we may have overestimated the
number of stars in the youngest two groups (and for them only at angular momenta near that of the
local standard of rest), we have probably not overestimated the number of stars in the older groups.

In this discussion we have neglected phase variations in the epicyclic angle.   However,
the epicyclic frequency is faster (about 40\% faster) than the angular rotation rate, so we expect the shearing in epicyclic
angle  takes place faster than in azimuthal angle.  

Equation \ref{eqn:theta} assumes
that stars were initially at the same azimuthal angle and had a similar angular momentum distribution.
Heating and migration could have taken place well after the birth of the group.
In this case the group would be less evenly distributed than estimated using its age and its current angular momentum 
distribution.
 If the abundance group originated in a star cluster that remained bound for a long time
before disrupting (e.g., \citealt{lamers06}) then the group would be less evenly distributed
than estimated here. However as a recently disrupted cluster 
should have a very narrow angular momentum distribution, more recent heating and migration rate would be required
to account for wide current eccentricity and angular momentum distributions.

The parent populations appear to be clumpy, however this could be due to sparse sampling.
Alternately phase wrapping due to shearing of azimuthal and epicyclic variations could
also cause clumping along this boundary \citep{minchev09}.
To estimate the parent population distributions we divide by a probability  that is
 sensitive to the eccentricity and angular momentum value near the forbidden boundary 
(as we can see from the sampling we used
in Figure \ref{fig:lprb}).  Along the forbidden region boundary, a small error in eccentricity or angular momentum 
could give a difference in  probability  of a factor of a few, and it is precisely in this region where
most of the stars are located because that is the only region where the probability of finding a star is high.
Errors in eccentricity and angular momentum measurements 
could cause the appearance of clumping near the forbidden boundary.
We have minimized this effect by taking the maximum probability within the estimated
errors for each data point. Nevertheless a small variation in a star's eccentricity and angular momentum
along this boundary causes a large change in probability and we should be careful when
interpreting structure in the parent populations.

We see from Figures \ref{fig:hist}, \ref{fig:probi}, and \ref{fig:probj}  that the youngest two groups  
have low eccentricity means and  dispersions.
The  estimated parent populations are large, greater than a million stars,
the large size arising because a small fraction of the thin disk stars were selected for study by \citet{bensby14}
and our correction for this selection increased the estimated number of parent stars.
Both groups contain weak tails in the distribution extending to higher eccentricity.
Only the 4.8 Gyr old grouping exhibits a tail toward higher angular momentum,
corresponding to stars coming from outside the Solar galactocentric radius.
It is difficult to determine whether the parent population distribution has a large angular momentum dispersion (width)
as the mean angular momentum values are near that of the Sun and low eccentricity regions above and below this value
lie in the forbidden region.
If there was a large low eccentricity population just interior to the Sun, then the higher eccentricity tails suggest
that the eccentricity width of the parent population is wider at lower angular momentum
than near $L \sim 1800$ \kmsp.
The estimated parent distributions suggest that most stars in the parent populations have not significantly migrated
(changed in angular momentum)
in the last 4-5 Gyr, though the tails in the parent distributions are significant.  Perhaps the same population that
migrated also increased in eccentricity dispersion and a skewed Gaussian model for migration might be preferred
(see Figure 3 by \citealt{bland10}).

The 7.1 year old group has a moderate width in its angular momentum distribution 
with a standard deviation of  200 \kmsp ~ and a mean of  $L \sim 1640$ \kmsp (and for
comparison to the other groups see  Table \ref{tab:parents}).
The mean angular momentum value is below that of the local standard of rest, 
and much of the parent population lies distant from the forbidden region, though
the parent population could extend to lower angular momentum ($L\lesssim 1500$ \kmsp) and 
low eccentricity ($e<0.2$), and into the forbidden region.
The tail of the distribution below $L=1300$ \kmsp ~ and $e>0.3$ suggests that the parent
population could contain low eccentricity stars below $L=1300$ \kmsp, as the eccentricity dispersion there is larger
than at the mean $L \approx 1640$ \kmsp. 
The parent angular momentum distribution (shown in
Figure \ref{fig:hist}) has  one strong major peak, similar to those of the two youngest groups.
In contrast the three oldest groups have much wider angular momentum distributions (also see the standard deviations listed
in Table \ref{tab:parents}).  The shape of the parent population angular momentum distribution for the 7.1 Gyr  old group 
suggests that many stars have not significantly migrated, however both width and fraction of stars in
the low angular momentum tail are higher at 7.1 Gyr than for the two younger groups.

The peaks in the parent populations of the three youngest groups  suggest that the bulk of their stars experienced
little migration within 7 Gyrs.  
Tails in the distributions imply that stars that have migrated in these groups have also increased in eccentricity dispersion.
However, a thin disk group that increased in angular momentum dispersion (due to migration) 
without increasing in eccentricity dispersion would not have stars present in the solar neighborhood unless
its mean angular momentum was near that of the LSR.    

As none of the peaks in the distributions (Figure \ref{fig:probj}) for the 3 older groups contain many stars we don't
attribute any significance to the individual peaks.
However, the oldest groups have both wide eccentricity and angular momentum parent distributions,
suggesting
that both heating and migration has taken place.

\begin{table*}  
\begin{minipage}{100mm}
 \caption[]{Estimated Properties of the Parent Populations \label{tab:parents}}
 \begin{tabular}{l r r  r r r r r r }
 \hline
 GID &  Age & $N_1$ & $N_2$ & $\langle L \rangle$ &$\sigma_L$ & $\langle e\rangle$ & $\sigma_e$\\
         &  (Gyr) &         &              & \multicolumn{2}{c}{(\kmsp) }           &        &\\
 \hline
5 & 4.0  & 275,550 & 3,786,181  & 1772 & 102 & 0.08 & 0.06\\
1 & 4.8  & 612,173 & 7,573,869  & 1754 & 167 & 0.12 & 0.06\\
3 & 7.1  & 466,881 & 4,993,675  & 1661 & 88 & 0.11 & 0.07\\
4 & 10.1 & 323,715 & 871,146    & 1511 & 266 & 0.25 & 0.11\\
2 & 10.2 & 315,132 & 1,105,883  & 1601 & 310 & 0.23 & 0.11\\
6 & 12.1 & 80,620  & 267,500    & 1679 & 282 & 0.23 & 0.15\\
\hline
\end{tabular}
 \\
GIC is the group number given by \citet{mitschang14} and their estimated age given in Gyr.
 $N_1$ is the estimated size of the parent population computed using equation \ref{eqn:N1} and taking
into account the probability of detecting an orbit in the solar neighborhood.
$N_2$ is the estimated size of the parent population computed using equation \ref{eqn:N2} and in addition
corrects the probability with an estimate for the selection function for the observed sample.
Mean eccentricity and angular momentum standard deviations are computed
from the derived parent populations (shown in Figure \ref{fig:probj}).
\end{minipage}
\end{table*}

\section{Summary and Discussion}

We summarize our primary findings here.  A discussion follows.
 
\begin{enumerate}[1.]
\item
 We find that stars in the 6 largest abundance groups by \citet{mitschang14} 
 tend to lie near a boundary in angular momentum vs eccentricity space 
where the probability is highest for a
star to be found in the Solar neighborhood, assuming a relaxed parent population evenly distributed
in azimuthal and epicyclic angles.  The stars that are most likely found
are those with orbital apocenter approximately equal to the Sun's galactocentric radius.
The bias has previously been described as a crossing time bias \citep{mayor77}.
\item
Using the probability for a star to be located in the Solar neighborhood (as a function
of eccentricity and angular momentum) and a crudely
estimated selection function for the sample, we estimate that
the parent populations of the  abundance groups 
range from  200,000 to a few million members.
\item
The two youngest groups lie nearest forbidden boundaries, implying that there could be a significant
population of group stars that cannot be seen in the Solar neighborhood.  
However the two youngest groups are the least likely to be evenly distributed azimuthally in the Galaxy and
by assuming an even distribution we may have over estimated the size of the parent populations by a factor of a few.
The angular momentum dispersions of the older groups imply that  the parent populations
are distributed at all azimuthal angles in the Galaxy and that we have not overestimated the sizes of their parent populations.
\item
Assuming that mean angular momentum is similar to that at birth, the width of the 
 parent populations of the thin disk groups suggest that the bulk of their stars experienced
little migration within 7 Gyrs.  
Tails in the distributions suggest that stars that have migrated in these groups have also increased in eccentricity dispersion.
In contrast,
the parent populations of the thick disk groups exhibit both wide angular momentum and eccentricity distributions
suggesting that both heating and radial migration has taken place.

\end{enumerate}

Here we assumed that eccentricity and inclination distributions are not correlated and have ignored
the vertical motions.  Using the vertical velocities it is possible to estimate
the inclination distribution of the parent populations.   Stars with high inclination are less likely
to be detected within 100pc of the Sun \citep{mayor77} and we have not taken this into account in our estimate
of the parent populations.  The numbers of stars in the older groups, with the highest vertical amplitudes, have been 
underestimated by a factor of a few due to this neglect.  

A large cluster may self pollute with supernova and so 
may not remain chemically homogeneous.  Consequently, single abundance populations are estimated 
to have sizes below $2 \times 10^5$ stars
(section 3.2 by \citealt{bland10}). 
The large sizes for the parent populations estimated here are a concern
as they are  above this limit.   One possibility is that each group may be comprised of similar but not identical fragments \citep{mitschang14}.  
Or the large groups may be part of a co-eval population composed of stars born nearly at the same {\it time}, and with 
similar abundances \citep{blanco15}, 
but  not necessarily all born in the same {\it place} (a co-eval but not necessarily co-natal parent population).
Alternatively the large parent population sizes could be attributed
to overestimation resulting from our assumption of an axisymmetric and mixed parent distribution. 
Smaller abundance groups were found by \citet{mitschang14} and these would
be consistent with the smaller parent sizes estimated for chemically homogenous populations.

We mention some uncertainties that affect this study.
\citet{bensby14} gave no selection description for their entire sample.
We crudely modeled the \citet{bensby14} sample distribution by comparing it with the GCS, however,  
the GCS sample itself is taken from two different magnitude limited source catalogs 
and is only complete to 40 pc \citep{nordstrom04}.
Future attempts to study parent populations of abundance groups
will be more robust if they are based on well characterized samples, 
and well characterized samples would allow  more robust estimates of parent populations.

In this study we used a Monte Carlo simulation technique to estimate the probability that
an orbit family would be detected in the Solar neighborhood.  We then used
this probability distribution and the stars in each group to estimate the parent population distributions.    
However, different distributions
for the parent populations could be assumed from the start and Monte Carlo simulations used to
predict the number and distribution of stars detected in the Solar neighborhood.  
This approach might alleviate some of the difficulties caused by the sparse sampling
resulting from the few stars in each group.

As did \citet{bensby14}, we adopted the Milky Way model by \citet{allen91}.
This study could be redone with different or updated Milky Way mass distributions to see how
the estimated parent populations are dependent upon the underlying assumed Galactic mass distribution.
Both accurate space motions and a good Milky Way mass model are needed to better estimate
the parent population distributions, particularly for stars near the forbidden boundary where
the probability is a  strong function of eccentricity and angular momentum.

We assumed a sharp edged spherical boundary at 100 pc from the Sun for the solar neighborhood sample.
However approximately 5\% of the stars from the 6 abundance groups are at larger distances.   
Future work could study the impact
of a selection function that depends on distance from the Sun.
Errors in measurement of eccentricities and angular momenta have been neglected from this study.
These too could be more accurately modeled.

Azimuthal structure in the probability distributions has been ignored in this study, however
the probability distributions could be sensitive to position with respect to the Galactic bar,
spiral arms and other dynamical structures such as the Galactic warp.

Future studies may detect variations in the orbital properties of the groups in different directions
allowing a study of azimuthal variations and correlations between orbital properties as a function
of distance from the Sun.  
As more stars are identified in a single group, it will be possible to determine
whether clumps in $e,L$ are real. Clumps along the region of high probability in $e,L$ space might arise
because of a non-uniform distribution in epicyclic amplitude (e.g., see \citealt{minchev09}).
The location of peaks in the distribution might depend on distance from the Sun, particularly
if the group is not well mixed in the Galaxy.
Detected structures would be exciting to study  with models of 
how groups evolve as they move in the Galaxy.

In summary, we were surprised by the large sizes of our estimated parent populations. 
The large sizes imply that large abundance 
groups found in the vicinity of the Sun are unlikely to be
co-natal populations unless they are unevenly distributed in the Galaxy. 
If the groups are not co-natal then they may not be comprised
of stars exactly the same age. 
The color magnitude diagram fits to the abundance groups 
were no worse than those of open clusters \citep{mitschang14}, 
suggesting that if there is an age spread in each group, it is not large.
However,
an age spread in the stars in the low metallicity groups might contribute to
the large eccentricity and angular momentum dispersions of these groups.
Likewise the higher metallicity groups may have lower angular momentum and eccentricity 
dispersions simply because they are comprised of younger and thin disk stars.
Despite these concerns, the increasingly large samples of stars
with accurate abundance measurements (e.g., \citealt{desilva15}) 
should be used to study 
groups of stars with similar abundances and may be used to probe mechanisms 
such as migration.
However, constraints on the dynamical evolution of stellar sub-populations 
will require larger and better characterized samples, samples that
extend away from  the solar neighborhood and comparisons between observed
and predicted distributions of many streams and groups.

\vskip 1.0 truein

We thank Segev Benzvi, Cameron Bell, 
Chelsea Jean, Eva Bodman, and Eric Mamajek for helpful discussions and correspondence.
BA and DBZ gratefully acknowledge the financial support of the Australian Research Council through Super Science Fellowship FS110200035 and Future Fellowship FT110100743, respectively.
This work was in part supported by NASA grant NNX13AI27G.
The authors would like to thank members of the Macquarie University and Australian Astronomical Observatory joint Galactic archaeology group for helpful discussions.  

{}

\end{document}